\documentclass[nofootinbib,twocolumn,prd,groupedaddress,preprintnumbers]{revtex4-1}
\usepackage{slashed}
\usepackage{bm}
\usepackage{graphicx}
\usepackage{color}
\usepackage{upgreek}
\usepackage{calligra}
\usepackage{amssymb}
\usepackage{mathrsfs}
\usepackage{amsmath}
\usepackage[normalem]{ulem}
\usepackage{bbm}
\usepackage{array}
\usepackage[utf8]{inputenc}
\usepackage[usenames,dvipsnames]{xcolor}
\RequirePackage[colorlinks=true,urlcolor=Magenta,anchorcolor=blue,citecolor=blue,filecolor=blue,
linkcolor=Magenta,menucolor=blue,linktocpage=true,pdfproducer=medialab]{hyperref}
\usepackage{catchfile} 
\newcommand{\getenv}[2][]{%
  \CatchFileEdef{\temp}{"|kpsewhich --var-value #2"}{}%
  \if\relax\detokenize{#1}\relax\temp\else\let#1\temp\fi}
\getenv[\USER]{USER}

\def\cG{\mathcal{G}}
\def\cg{\text{\calligra g}}
\def\cY{\mathcal{Y}}
\def\cO{\mathcal{O}}
\def\VV{\mathscr{V}}
\def\LL{\mathscr{L}}
\def\nn{\nonumber}
\DeclareMathOperator{\diag}{\texttt{diag}}

\DeclareMathOperator{\derp}{\partial}

\newcommand{\ov}[1]{\overline{#1}}
\newcommand{\vev}[1]{\langle#1\rangle}
\def\be{\begin{equation}}
\def\ee{\end{equation}}
\newcommand{\TeV}{\;\text{TeV}}
\newcommand{\GeV}{\;\text{GeV}}
\newcommand{\MeV}{\;\text{MeV}}
\newcommand{\keV}{\;\text{keV}}
\newcommand{\eV}{\;\text{eV}}
\newcommand{\meV}{\;\text{meV}}
\newcommand{\mueV}{\;\text{$\mu$eV}}
\newcommand{\m}{\;\text{m}}
\newcommand{\mm}{\;\text{mm}}
\newcommand\UFN{U(1)_{\text{FN}}}
\newcommand\UPQ{U(1)_{\text{PQ}}}
\newcommand{\blue}[1]{\color{blue} #1 \color{black} }

\begin{document}

\title{\blue{The Minimal Flavour Violating Axion}}

\author{F.~Arias-Arag\'on}
\email{fernando.arias@uam.es}
\affiliation{\vspace{1mm} 
Departamento de F\'isica Te\'orica and Instituto de F\'isica Te\'orica, IFT-UAM/CSIC, Universidad Aut\'onoma de Madrid, Cantoblanco, 28049, Madrid, Spain}
\author{L.~Merlo}
\email{luca.merlo@uam.es}
\affiliation{\vspace{1mm} 
Departamento de F\'isica Te\'orica and Instituto de F\'isica Te\'orica, IFT-UAM/CSIC, Universidad Aut\'onoma de Madrid, Cantoblanco, 28049, Madrid, Spain}

\begin{abstract}
The solution to the Strong CP problem is analysed within the Minimal Flavour Violation (MFV) context. An Abelian factor of the complete flavour symmetry of the fermionic kinetic terms may play the role of the Peccei-Quinn symmetry in traditional axion models. Its spontaneous breaking, due to the addition of a complex scalar field to the Standard Model scalar spectrum, generates the MFV axion, which may redefine away the QCD theta parameter. It differs from the traditional QCD axion for its couplings that are governed by the fermion charges under the axial Abelian symmetry. It is also distinct from the so-called Axiflavon, as the MFV axion does not describe flavour violation, while it does induce flavour non-universality effects. The MFV axion phenomenology is discussed considering astrophysical, collider and flavour data. The alternative Axion-like-particle possibility is also investigated.
\end{abstract}

\preprint{FTUAM-17-17}
\preprint{IFT-UAM/CSIC-17-086}

\maketitle

%
%
\section{Introduction}

The search for an explanation of the fermion mass heterogeneity and of the different mixing schemes in the quark and lepton sectors underwent to a strong activity in the last 40 years. Rejecting the anthropic creed, the best strategy is to add a flavour symmetry to the Standard Model (SM) gauge group: this symmetry rules the fermion couplings, explaining the observed flavour puzzle and determining the amount of flavour violation in the theory.

The simplest possibility consists in the Abelian continuous $\UFN$ flavour symmetry, dubbed Froggatt-Nielsen (FN), first considered in Ref.~\cite{Froggatt:1978nt}. Fermions may transform under the FN group and the Yukawa operators are invariant under $\UFN$ transformations only introducing powers of an additional real scalar field $\phi$, with a non-trivial $U(1)_{FN}$ charge. Yukawa terms turn out to be non-renormalisable and then suppressed by suitable powers of the cut-off scale $\Lambda_F$ of the theory. Once the scalar field $\phi$, typically called flavon, develops a vacuum expectation value (VEV) the flavour symmetry is spontaneously broken and the Yukawa matrices can be written in terms of powers of $\vev{\phi}/\Lambda_F$. Fermion mass hierarchies and mixing angles can then be explained by an appropriate choice of the FN charges~\cite{Altarelli:2000fu,Altarelli:2002sg,Chankowski:2005qp,Buchmuller:2011tm,Altarelli:2012ia,Bergstrom:2014owa}. The large number of free parameters, one for each entry of the Yukawa matrices, has the drawback of lowering the predictive power of the model: any value of fermion masses and mixings can indeed be reproduced. Moreover, in order not to spoil the so good agreement of the SM predictions on flavour observables with the experimental data, the scale $\Lambda_F$ is constrained to be much larger than the electroweak (EW) scale.

More economical models in terms of number of parameters have been proposed only subsequently, based on non-Abelian discrete or continuous symmetries. The first class of theories exhibited very predictive mass textures~\cite{Ma:2001dn,Babu:2002dz,Altarelli:2005yp,Altarelli:2005yx,Feruglio:2007uu,Bazzocchi:2009pv,Bazzocchi:2009da,Altarelli:2009gn,Toorop:2010yh,Altarelli:2010gt,Toorop:2010yh,Varzielas:2010mp,Toorop:2011jn,Grimus:2011fk,deAdelhartToorop:2011re,King:2011ab,Altarelli:2012ss,Bazzocchi:2012st,King:2013eh} and provided a certain protection from flavour violating processes~\cite{Feruglio:2008ht,Feruglio:2009iu,Lin:2009sq,Feruglio:2009hu,Ishimori:2010au,Toorop:2010ex,Toorop:2010kt,Merlo:2011hw,Altarelli:2012bn}. However, the 2011 discovery of a non-vanishing, and relatively large, leptonic reactor angle~\cite{Abe:2011sj,Adamson:2011qu,Abe:2011fz,An:2012eh,Ahn:2012nd} has raised strong doubts on the use of non-Abelian discrete models, whose most common prediction was a vanishing reactor angle.

Non-Abelian continuous symmetries, instead, have shown to be effective to describe the SM flavour puzzle and to keep well under control flavour violating contributions from new physics. The most known context is the so-called Minimal Flavour Violation (MFV) that consists in the simple ansatz~\cite{Chivukula:1987py} that any source of flavour and CP violation in any theory Beyond the SM (BSM) is the one in the SM, i.e. the Yukawa couplings. This concept can be technically formulated in terms of the flavour symmetry arising in the considered Lagrangian in the limit of vanishing Yukawa couplings: the flavour group is a product of a $U(3)$ factor for each field species in the given spectrum; in the SM case it is $U(3)^5$~\cite{DAmbrosio:2002vsn}\footnote{Once considering the BSM extension with three right-handed neutrinos, which allows for an explanation of the active neutrino masses, the flavour group is $U(3)^6$~\cite{Cirigliano:2005ck,Davidson:2006bd,Alonso:2011jd}. This scenario is however not predictive and a reduction of the symmetry is required. See Ref.~\cite{Dinh:2017smk} for a recent update.}. Yukawa terms are invariant under this flavour symmetry only promoting the Yukawa matrices to be fields transforming non-trivially under the non-Abelian part of $U(3)^5$. In the original formulation~\cite{DAmbrosio:2002vsn}, the Yukawa spurions are non-dynamical fields, of vanishing mass dimension, that acquire specific background values, which exactly reproduce the measured fermion masses and mixing angles. Any non-renormalisable operator constructed with the SM fields is, eventually, made flavour invariant by suitably inserting the Yukawa spurions: once expliciting the background values, the strength of the effects induced by these operators in flavour violating observables is suppressed by specific combinations of fermion masses, mixing angles and CP violating phases. In consequence, once considering the constraints from flavour data, the scale $\Lambda_F$ of the new physics originating the non-renormalisable operators, can be of the order of a few TeV~\cite{DAmbrosio:2002vsn,Cirigliano:2005ck,Cirigliano:2006su,Davidson:2006bd,Grinstein:2006cg,Hurth:2008jc,Kagan:2009bn,Gavela:2009cd,Grinstein:2010ve,Feldmann:2010yp,Guadagnoli:2011id,Alonso:2011jd,Buras:2011zb,Buras:2011wi,Alonso:2012jc,Isidori:2012ts,Lopez-Honorez:2013wla,Bishara:2015mha,Lee:2015qra,Feldmann:2016hvo,Alonso:2016onw,Dinh:2017smk,Merlo:2018rin}, instead of hundreds of TeV in the generic case~\cite{Isidori:2010kg}.

Although the MFV is a very predictive context, fermion masses and mixings are only described but not explained: no justification of the Yukawa background values is provided. Steps forward a completion of the MFV framework have been taken in Refs.~\cite{Alonso:2011yg,Alonso:2012fy,Alonso:2013mca,Alonso:2013nca} (see also Refs.~\cite{Anselm:1996jm,Barbieri:1999km,Berezhiani:2001mh,Feldmann:2009dc,Nardi:2011st}): the Yukawa spurions have been promoted to dynamical scalar fields and the corresponding scalar potential has been investigated. This analysis showed interesting, even if not conclusive, results: a minimum of the potential describes non-vanishing masses for the heavier charged fermions, two non-vanishing neutrino masses, almost no mixing in the quark sector, one maximal lepton mixing associated to a maximal Majorana phase, when considering the SM fermion spectrum extended by the addition of three right handed neutrinos.

Once a continuous symmetry is spontaneously broken, Goldstone Bosons (GBs) are generated. This possibility is typically avoided within the MFV context gauging (part of) the symmetry~\cite{Grinstein:2010ve,Feldmann:2010yp,Guadagnoli:2011id,Buras:2011zb,Buras:2011wi,Feldmann:2016hvo,Alonso:2016onw}. On the other side, surviving GBs may represent the key ingredients to deal with other open problems in the SM. The focus in this letter will be on the Strong CP problem and it will be shown that a GB arising from the spontaneous breaking of an Abelian term of the MFV symmetry $U(5)^3$ can be an axion. 

The solution of the Strong CP problem proposed here in the MFV context follows the lines of the traditional QCD axion~\cite{Peccei:1977hh,Wilczek:1977pj,Weinberg:1977ma}: the $U(1)$ factor that originates the MFV axion is not vectorial and it is explicitly broken by the colour anomalies; the so-called theta-parameter,
\be
\LL_\text{QCD}\supset\dfrac{\alpha_s}{8\pi}\theta_{QCD}G^{a\mu\nu}\widetilde{G}^a_{\mu\nu}
\ee
with $\widetilde{G}^a_{\mu\nu}\equiv \frac{1}{2}\epsilon_{\mu\nu\rho\sigma} G^{a\rho\sigma}$ and $\epsilon_{\mu\nu\rho\sigma}$ the totally antisymmetric tensor such that $\epsilon_{1230}=1$, can be redefined away by a shift symmetry transformation; exactly as for the QCD axion (see for example Ref.~\cite{diCortona:2015ldu}), the minimum of the MFV axion potential is in zero and in consequence the QCD CP violating term exactly vanishes.

The MFV axion differs from the traditional QCD axion and from the so-called invisible axions~\cite{Kim:1979if,Shifman:1979if,Zhitnitsky:1980tq,Dine:1981rt} as the transformation properties under the axial $U(1)$ factor are determined by the flavour structure of the SM fermions. Its associated phenomenology will be discussed in astrophysics, collider searches and in flavour observables, mainly focussing on meson decays.

The MFV axion manifests different signatures even respect with the so-called Axiflavon or Flaxion, recently presented in Refs.~\cite{Ema:2016ops,Calibbi:2016hwq}, based on the pioneering paper in Ref.~\cite{Wilczek:1982rv}. The Axiflavon is the GB arising from the spontaneous breaking of the flavour $\UFN$ symmetry in the FN mechanism and its distinctive feature resides in its flavour violating couplings. On the contrary, the MFV axion presents flavour conserving couplings, but violating the flavour universality. The predictions for meson decays are therefore different. 

The rest of the letter is structured as follows. Few selected features of the MFV context are reported in Sect.~\ref{Sect:MFV}. The MFV axion is presented in Sect.~\ref{Sect:MFVAxion}. Its phenomenology is discussed in Sect.~\ref{Sect:Pheno} together with a comparison with the Axiflavon. Conclusive remarks can be found in Sect.~\ref{Sect:Conclusions}. 
%
%
\section{The Minimal Flavour Violation Revisited}
\label{Sect:MFV}

According to the modern realisation of MFV~\cite{DAmbrosio:2002vsn,Cirigliano:2005ck,Davidson:2006bd,Alonso:2011jd,Dinh:2017smk}, the SM fermionic kinetic terms exhibit a $U(3)^5$ flavour symmetry that can be decomposed into the product of an Abelian and a non-Abelian factor, $\cG_F^\text{NA}\times \cG_F^\text{A}$ where
\be
\hspace{-0.3cm}
\begin{aligned}
\cG_F^\text{NA}&\equiv SU(3)_{q_L}\times SU(3)_{u_R}\times SU(3)_{d_R}\times SU(3)_{\ell_L}\times SU(3)_{e_R}\\
\cG_F^\text{A}&\equiv U(1)_B\times U(1)_L\times U(1)_Y\times \UPQ\times U(1)_{e_R}\,.
\end{aligned}
\ee
In the previous expressions, $q_L$ and $\ell_L$ stand for the quark and lepton $SU(2)_L$ doublets, while $u_R$, $d_R$ and $e_R$ for the quark and lepton singlets, each of them transforming as a triplet of the corresponding symmetry group; $B$ and $L$ refer to the Baryon and Lepton numbers, $Y$ to the Hypercharge, PQ to the PQ symmetry, while the last Abelian symmetry factor corresponds to rotations on only the $e_R$ fields.

In order to guarantee the invariance under this flavour symmetry of the entire SM Lagrangian, the Yukawa matrices $Y_i$ are promoted to spurion fields $\cY_i$ transforming under $\cG_F^\text{NA}$ as
\be
\begin{gathered}
\cY_u\sim({\bf3},\,{\bf\ov3},\,1,\,1,\,1)\qquad
\cY_d\sim({\bf3},\,1,\,{\bf\ov3},\,1,\,1)\\
\cY_e\sim(1,\,1,\,1,\,{\bf3},\,{\bf\ov3})\,.
\end{gathered}
\ee
The Yukawa spurions then acquire background values, which explicitly break $\cG_F^\text{NA}$ and describe fermion masses and mixings. A free choice for these values is the ensamble of fermion masses and mixing angles, that can be written as follows:
\be
\begin{aligned}
\vev{\cY_u}&=c_tV^\dag\diag\left(\dfrac{m_u}{m_t},\,\dfrac{m_c}{m_t},\,1\right)\\
\vev{\cY_d}&=c_b\diag\left(\dfrac{m_d}{m_b},\,\dfrac{m_s}{m_b},\,1\right)\\
\vev{\cY_e}&=c_\tau\diag\left(\dfrac{m_e}{m_\tau},\,\dfrac{m_\mu}{m_\tau},\,1\right)\,.
\end{aligned}
\label{YukawaBackgrounds}
\ee
where $m_i$ are the fermion masses, $V$ is the CKM mixing matrix and $c_i$ are global numerical factors, not larger than 1\footnote{Considering values of $c_i$ larger than 1 implies that multiple products of Yukawa spurions would be more relevant than the single spurions themselves, and then they should be treated in a non-perturbative approach as discussed in Ref.~\cite{Kagan:2009bn}.}. Neutrino masses and the PMNS mixing matrix, although their introduction through the Seesaw mechanism is straightforward, will not be tackled in this letter as the focus is the MFV axion and the solution of the strong CP problem.

Refs.~\cite{Alonso:2011yg,Alonso:2012fy,Alonso:2013mca,Alonso:2013nca} showed how the non-Abelian symmetry $\cG^{NA}_F$ deals exclusively with the explanation of the inter-generation hierarchies, but cannot fix the overall coefficients $c_i$. On the other side, the hierarchies $m_b/m_t$ and $m_\tau/m_t$, that correspond to $c_b/c_t$ and $c_\tau/c_t$ in the previous expressions, can be elegantly explain {\it \`a la} FN mechanism: one of the Abelian factors of the whole flavour symmetry can be taken as a true symmetry of the Lagrangian; charges can be chosen such that the only terms invariant under this Abelian factor are the ones of the up-type quarks, while those describing down-type quarks and charged leptons are initially forbidden. The latter terms originate only at the non-renormalisable level, after the addition in the scalar spectrum of a new field, the flavon $\Phi$, transforming under this Abelian factor, which re-establishes the invariance under the symmetry. 

The minimality criterium in terms of field content identifies $\UPQ$ as the only candidate\footnote{Ref.~\cite{Albrecht:2010xh} discussed a similar context, where the non-Abelian terms of the flavour symmetry are gauged and two distinct Abelian factors are considered to explain the ratio between the third family quark masses. The spectrum and flavour gauge symmetries lead to a different phenomenology with respect to the one discussed in Refs.~\cite{Grinstein:2010ve,Buras:2011wi} and here.}, among the Abelian factors in $\cG_F^\text{A}$, to explain the intra-generation hierarchies: Baryon and Lepton numbers and Hypercharge are fixed by definition; $U(1)_{e_R}$ does not affects down-type quarks and therefore would only explain the ratio $m_\tau/m_t$; although a double FN mechanism could be possible, taking both $\UPQ$ and $U(1)_{e_R}$ as true symmetries of the Lagrangian would lead to the introduction of two flavons, increasing the complexity of the model. Instead, as all the fermions potentially transform under $\UPQ$, this choice allows to explain both the rations $m_b/m_t$ and $m_\tau/m_t$ by introducing a single scalar field. This is the strategy adopted in the following. 

Without specifying, for sake of generality, the fermionic $\UPQ$ charge assignment, the Yukawa Lagrangian reads as
\begin{align}
\hspace{-0.3cm}
\LL_Y=&-\left(\dfrac{\Phi}{\Lambda_\Phi}\right)^{x_u-x_q}\ov{q}_L\widetilde{H}\cY_u u_R-\left(\dfrac{\Phi}{\Lambda_\Phi}\right)^{x_d-x_q}\ov{q}_LH\cY_d d_R+\nn\\
&-\left(\dfrac{\Phi}{\Lambda_\Phi}\right)^{x_e-x_\ell}\ov{\ell}_LH\cY_e e_R\,,
\label{YukawaAxionLagrangian}
\end{align}
where $H$ is the Higgs $SU(2)_L$ doublet, $\widetilde{H}=i\sigma_2H^*$, and $x_i$ are the PQ charges of the $i$ field\footnote{The mixed use of a PQ flavon $\Phi$ and of the Yukawa spurions $\cY_i$ may be puzzling. Indeed, at this level of the discussion, it is equivalent introducing a dynamical flavon $\Phi$ or treating its effects via a PQ spurion (see Ref.~\cite{Alonso:2011jd} for the latter case). Similarly, the Yukawa spurions may be promoted to be dynamical fields (see Refs.~\cite{Alonso:2011yg,Alonso:2012fy,Alonso:2013mca,Alonso:2013nca}). The discussion that follows and the results presented in this section are not affected by this choice. Instead, the necessity to describe the breaking of the PQ symmetry through a dynamical flavon resides in the origin of the MFV axion, as it will be explained in the Sect.~\ref{Sect:MFVAxion}.}. In this expression, the charge of the flavon $\Phi$ has been fixed to $-1$, without loss of generality.

Once this flavon develops a VEV, $\vev{\Phi}\equiv v_\Phi$, and the Yukawa spurions acquire their background values, the Yukawa interactions read
\be
\begin{gathered}
Y_u=\epsilon^{x_u-x_q}\vev{\cY_u}\qquad Y_d=\epsilon^{x_d-x_q}\vev{\cY_d}\\
Y_e=\epsilon^{x_e-x_\ell}\vev{\cY_u}\,,
\end{gathered}
\label{FinalYukawas}
\ee
where $\epsilon\equiv v_\Phi/\sqrt2\Lambda_\Phi$. It follows that the ratios between the third generation fermions are governed by the specific fermion PQ charge assignment: 
\be
\begin{aligned}
m_b/m_t&\simeq \epsilon^{x_d-x_u}\\
m_\tau/m_t&\simeq \epsilon^{x_e-x_\ell-x_u+x_q}\,,
\end{aligned}
\ee 
where the ratios of the $c_i$ factors have been omitted as they are expected to be of the same order. The fact that the top mass is given by $c_t v/\sqrt2$ implies that $c_t\simeq1$ and it results in selecting as the simplest choice
\be
x_q=x_u=0
\label{PQChargesChoice1}
\ee
for the quark doublets and up-quark singlets PQ charges. In consequence, the other charges must undergo the following relations,
\be
\begin{gathered}
x_d\simeq\log_\epsilon(m_b/m_t)\\
x_e-x_\ell\simeq\log_\epsilon(m_\tau/m_t)
\end{gathered}
\label{PQChargesChoice2}
\ee
where $\epsilon$ is still an unknown quantity at this level. An exact value for this parameter depends on the specific ultraviolet theory that originates the low-energy Lagrangian in Eq.~(\ref{YukawaAxionLagrangian}). An estimation of the range of value it may acquire takes into consideration that $\epsilon$ should remain in the perturbative regime and that the value of $v_\Phi$ is expected to be not so much smaller than $\Lambda_\Phi$ (without a dynamical mechanism to explain it). In this letter, $\epsilon$ is taken in the interval $[0.01,\,0.3]$, consistently with previous studies on FN models~\cite{Altarelli:2012ia,Bergstrom:2014owa}. The logarithm in Eq.~(\ref{PQChargesChoice2}) softens the dependence on the exact value of $\epsilon$: for $\epsilon$ inside its preferred interval, $x_d$ and $x_e-x_\ell$ are found in the range $[1,\,4]$; to fix a reference value that will be used in the phenomenological analysis, 
\be
x_d=3\,,\qquad\qquad
x_e-x_\ell=3\,,
\label{PQChargesChoice3}
\ee
corresponding to $\epsilon\sim0.23$, i.e. the Cabibbo angle.   

Charged fermion masses and quark mixings do not help further to break the flat direction $x_e-x_\ell$. On the other side, neutrinos masses introduce an additional condition: describing neutrino mass terms via the Weinberg operator~\cite{Weinberg:1979sa}, invariance under $\UPQ$ implies that this operator is written with $2x_\ell$ insertions of the flavon $\Phi$,
\be
\LL_5=\left(\dfrac{\Phi}{\Lambda_\Phi}\right)^{2x_\ell}\times\dfrac{\left(\ov{\ell_L^c}\widetilde{H}^*\right)\cg_\nu \left(\widetilde H^\dag\ell_L\right)}{\Lambda_L}\,,
\ee
where $\Lambda_L$ is the scale of lepton number violation and $\cg_\nu$ is the spurion field, transforming as $(1,\,1,\,1,\,\ov6,\,1)$ under $\cG_F^\text{NA}$, whose background value $\vev{\,\cg_\nu}\equiv g_\nu$ contains the information of the neutrino mass eigenvalues and the PMNS mixing matrix (see Ref.~\cite{Cirigliano:2005ck,Dinh:2017smk} for details). Larger values of $x_\ell$ implies that a lower scale $\Lambda_L$ is sufficient to explain the lightness of the active neutrinos. Moreover, requiring that the eigenvalues of $g_\nu$ are not larger than 1 (as for $c_i$ in Eq.~(\ref{YukawaBackgrounds})), an upper bound on $\Lambda_L$ can be identified:
\be
\Lambda_L\simeq\dfrac{v^2}{2}\dfrac{g_\nu \,\epsilon^{2x_\ell}}{\sqrt{\Delta m^2_\text{atm}}}\lesssim 6\times 10^{14}\GeV\times \epsilon^{2x_\ell}\,,
\ee
where $\Delta m^2_\text{atm}$ is the atmospheric neutrino mass squared difference. Ref.~\cite{Dinh:2017smk} (see Fig.~1) shows that, for $x_\ell=0$, the present data on the $\mu\to e$ conversion in golden nuclei largely excludes the parameter space for this model. A straightforward computation for $x_\ell=2$ and $\epsilon=0.23$ easily reveals that the parameter space is practically excluded. This suggests that only two values, $x_\ell=0$ and $x_\ell=1$, should be considered in the phenomenological analysis that follows. Summarising, two scenarios will be studied\footnote{The stability of a generic choice for these charges under the renormalisation group evolution has been discussed in Ref.~\cite{Choi:2017gpf}, specially considering the impact on axion couplings, which will be the subject of the next section. These effects could be relevant if the axion scale $f_a$ is relatively small, while for the values considered here they can be neglected.}:
\begin{align}
\text{S0:}\qquad &x_q=0=x_u=x_\ell\,,\qquad x_d=3=x_e
\label{PQChargesChoice}\\
\text{S1:}\qquad &x_q=0=x_u\,,\quad x_\ell=1\,, \quad x_d=3\,, \quad x_e=4\,.\nn
\end{align}

%
%

\section{The MFV Axion}
\label{Sect:MFVAxion}

The origin of an axion in this context is associated to the PQ flavon: if $\Phi$ is a complex scalar field, then it contains two degrees of freedom, which in the polar coordinates can be expressed as follows: in the PQ broken phase,
\be
\Phi=\dfrac{\rho+v_\Phi}{\sqrt2}e^{ia/v_\Phi}\,,
\ee
where $\rho$ is the radial component and $a$, which can be identified with the axion, is the the angular one. 

The full scalar potential of the model presents three distinct parts: 
\be
\VV=-\mu^2|H|^2+\lambda|H|^4-\mu_\Phi^2|\Phi|^2+\lambda_\Phi|\Phi|^4+\lambda_{H\Phi}|H|^2|\Phi|^2\,.
\ee
In a part of the parameter space, the pure $\Phi$-dependent scalar potential has a minimum corresponding to a non-vanishing VEV for $\Phi$, $v_\Phi^2=\mu_\Phi^2/\lambda_\Phi$. 

If $v_\Phi$ is much larger than the EW scale, this may represent a problem for the EW symmetry breaking (EWSB) mechanism: indeed the quartic $|H|^2|\Phi|^2$ coupling would contribute to the quadratic term of the pure $H$-dependent potential,
\be
\mu^2 \to \mu^{\prime2}\equiv \mu^2-\lambda_{H\Phi}v_\Phi^2\,.
\ee
If no {\it ad hoc} cancellation between these two terms is invoked and for arbitrary values of $\lambda_{H\Phi}$, the new mass parameter $\mu'$ resides at the same, large scale of $v_\Phi$. In order to reproduce the expected value of the EW VEV, $v\equiv 245\GeV$ fixed through the $W$ gauge boson mass, it is then necessary to invoke a large value of the Higgs quartic coupling $\lambda$, describing in this way a strongly interacting scenario with a non-linearly realised EWSB mechanism. This is an intriguing possibility, especially considering the recent interest in non-SM descriptions of the Higgs sector, such as composite Higgs models~\cite{Kaplan:1983fs,Kaplan:1983sm,Banks:1984gj,Agashe:2004rs,Gripaios:2009pe,Mrazek:2011iu,Alonso:2014wta,Panico:2015jxa,Hierro:2015nna}, dilaton models~\cite{Halyo:1991pc,Goldberger:2008zz,Vecchi:2010gj,Matsuzaki:2012mk,Chacko:2012sy,Chacko:2012vm,Bellazzini:2012vz,Hernandez-Leon:2017kea}, or general effective Lagrangians~\cite{Feruglio:1992wf,Contino:2010mh,Alonso:2012px,Alonso:2012pz,Buchalla:2013rka,Brivio:2013pma,Brivio:2014pfa,Gavela:2014vra,Gavela:2014uta,Brivio:2015kia,Gavela:2016bzc,Alonso:2016btr,Eboli:2016kko,Brivio:2016fzo,Alonso:2016oah,deFlorian:2016spz,Merlo:2016prs,Kozow:2019txg}. 

On the other side, if $v_\Phi$ is close to the EW scale, then no tuning is required and the EWSB mechanism would work as in the SM (see for example Refs.~\cite{Merlo:2017sun,Alonso-Gonzalez:2018vpc}). In the phenomenological section, both the cases will be discussed.

In this letter, $v_\Phi$ will be considered sufficiently large to consider the radial component $\rho$ as an heavy degree of freedom: it can be safely integrated out from the low-energy Lagrangian, leaving the axion $a$ as the unique light degree of freedom of $\Phi$. The low-energy Lagrangian of the model can therefore be written as the sum of distinct terms:
\be
\begin{aligned}
\LL=&\LL^\text{SM}_{Kin}+\dfrac{1}{2}\derp_\mu a\derp^\mu a+\mu^{\prime2}|H|^2-\lambda|H|^4+\\
&-e^{i(x_u-x_q)a/v_\Phi}\ov{q}_L\widetilde{H}Y_u u_R-e^{i(x_d-x_q)a/v_\Phi}\ov{q}_LHY_d d_R+\\
&-e^{i(x_e-x_\ell)a/v_\Phi}\ov{\ell}_LHY_e e_R+c^{(5)}_\nu e^{2i\,x_\ell\, a/v_\Phi}\left(\ov{\ell_L^c}\widetilde{H}^*\right)\left(\widetilde H^\dag\ell_L\right)\\
&+\dfrac{\alpha_s}{8\pi}\theta_{QCD}G^{a\mu\nu}\widetilde{G}^a_{\mu\nu}\,.
\label{InitialLagrangianLE}
\end{aligned}
\ee
where
\be
c^{(5)}_\nu\equiv\epsilon^{2x_\ell}\dfrac{\vev{\,\cg_\nu}}{\Lambda_L}=-\dfrac{2}{v^2}M_\nu\,.
\ee
Some comments are in order. The Yukawa matrices are the ones defined in Eq.~(\ref{FinalYukawas}) and $c^{(5)}_\nu$ is the coefficient of the Weinbeng operator and $m_\nu$ is the neutrino mass matrix in the flavour basis. The specific choice of the PQ charges in Eq.~(\ref{PQChargesChoice}) has not been implemented yet, to keep general the discussion. 

It is straightforward to rewrite the Lagrangian in the basis where the axion-fermion couplings are derivative. The resulting Lagrangian consists of the SM Lagrangian modified by the addition of interactions with the axion that read
\be
\delta\LL=\dfrac{1}{2}\derp_\mu a\derp^\mu a-c_{a\psi}\dfrac{\derp_\mu a}{2v_\Phi}\ov{\psi}\gamma^\mu\gamma_5\psi-M_\nu e^{2i\,x_\ell\, a/v_\Phi}\ov{\nu^c_L}\nu_L
\,,
\label{PhysicalAxionCouplingsFermions}
\ee
where $\psi=\{u,\,d,\,e\}$ and the coefficients are 
\be
\begin{aligned}
c_{au}=&x_q-x_u\\
c_{ad}=&x_q-x_d\\
c_{ae}=&x_\ell-x_e\,.
\end{aligned}
\label{eq:coefficientsFermions}
\ee

At the quantum level, the derivative of the axial current is non-vanishing, giving rise to the following effective axion-gauge boson couplings: in the physics basis for the gauge bosons
\be
\begin{aligned}
\delta\LL_\text{eff}\supset&
-\dfrac{\alpha_s}{8\pi}\, c_{agg}\dfrac{a}{v_\Phi} G^{a\mu\nu}\widetilde{G}^a_{\mu\nu}
-\dfrac{\alpha_{em}}{8\pi}\,c_{a\gamma\gamma}\dfrac{a}{v_\Phi}F^{\mu\nu}\widetilde{F}_{\mu\nu}+
\\
&
-\dfrac{\alpha_{em}}{8\pi}\,c_{aZZ}\dfrac{a}{v_\Phi}Z^{\mu\nu}\widetilde{Z}_{\mu\nu}
-\dfrac{\alpha_{em}}{8\pi}\,c_{a\gamma Z}\dfrac{a}{v_\Phi}F^{\mu\nu}\widetilde{Z}_{\mu\nu}+\\
&
-\dfrac{\alpha_{em}}{8\pi}\,c_{aWW}\dfrac{a}{v_\Phi}W^{+\mu\nu}\widetilde{W}^{-}_{\mu\nu}\,,
\end{aligned}
\label{PhysicalAxionCouplingsGaugeBosons}
\ee
where $X_{\mu\nu} = \partial_\mu X_\nu - \partial_\nu X_\mu$, and
\be
\begin{aligned}
c_{agg}=&3(c_{au}+c_{ad})\\
c_{a\gamma\gamma}=&2(4c_{au}+c_{ad}+3c_{ae})\\
c_{aZZ}=&\dfrac{t^2_\theta}{4}\left(17c_{au}+5c_{ad}+15c_{ae}\right)+\dfrac{3}{4t^2_\theta}\left(3(c_{au}+c_{ad})+c_{ae}\right)\\
c_{a\gamma Z}=&\dfrac{t_\theta}{4}\left(17c_{au}+5c_{ad}+15c_{ae}\right)-\dfrac{3}{4t_\theta}\left(3(c_{au}+c_{ad})+c_{ae}\right)\\
c_{aWW}=&\dfrac{3}{2s^2_\theta}\left(3\left(c_{au}+c_{ad}\right)+c_{ae}\right)\,,
\end{aligned}
\label{eq:coefficientsGaugeBosons}
\ee
with for shortness $t_\theta\equiv\tan\theta_W$, $s_\theta\equiv\sin\theta_W$ and $s_{2\theta}\equiv\sin2\theta_W$, being $\theta_W$ the Weinberg angle. The coefficients of the anomalous terms contain the contributions from all the fermions with a non-vanishing PQ charge.

As anticipated in the introduction, the MFV axion solves the Strong CP problem in exactly the same way as the traditional QCD axion: the $\theta_{QCD}$ parameter can be absorbed by a shift transformation of the axion. The only condition that must be satisfied is that $c_{agg}\neq0$, which is consistent with Eq.~(\ref{PQChargesChoice}), that explain the top Yukawa coupling of order 1 and the smallness of the bottom mass with respect to the top mass.

Table~\ref{tab:pheno} reports the values of the $c_{ai}$ coefficients of the axion couplings to fermions and gauge field strengths in the physical basis for the two scenarios described in Eq.~(\ref{PQChargesChoice}). As the coefficients in Eq.~\eqref{eq:coefficientsGaugeBosons} depend only on charge differences, the values for the anomaly couplings are the same values for both the scenarios. Of particular interest is that the ratio between the axion coupling to photons and that to gluons, which is typically a free parameter~\cite{Giudice:2012zp,Redi:2012ad,Redi:2016esr,DiLuzio:2016sbl,Farina:2016tgd,Coy:2017yex}, is exactly fixed to $8/3$, as in the original DFSZ invisible axion model. 

\begin{table}[h!]
\begin{center}
\begin{tabular}{c||c|c||c|c|c|c|c|c|c|c|}
 & $x_\ell$ & $x_e$ & $c_{au}$ & $c_{ad}$ & $c_{ae}$ &$c_{agg}$ & $c_{a\gamma\gamma}$ & $c_{aZZ}$ & $c_{a\gamma Z}$ &  $c_{a W W}$\\
\hline
$S0$ & $0$ & $3$ & $0$ & $-3$ & $-3$ & $-9$ & $-24$ & $-35.8$ & $8.8$ & $-81$\\
$S1$ & $1$ & $4$ & $0$ & $-3$ & $-3$ & $-9$ & $-24$ & $-35.8$ & $8.8$ & $-81$\\
\hline
\end{tabular}
\caption{\em Values of the coefficients of the axion couplings to fermions and gauge boson field strengths in the physical basis for the two scenarios identified in Eq.~(\ref{PQChargesChoice}), where the normalisation is defined in Eqs.~(\ref{PhysicalAxionCouplingsFermions}) and (\ref{PhysicalAxionCouplingsGaugeBosons}).}
\label{tab:pheno}
\end{center}
\end{table}

Notice that the common notation adopted in the literature makes use of effective couplings that can be expressed in terms of the $c_{ai}$ coefficients as follows:
\be
\begin{aligned}
g_{agg}&\equiv \dfrac{\alpha_{s}}{2\pi}\,\frac{c_{agg}}{v_\Phi}\equiv\dfrac{\alpha_{s}}{2\pi}\dfrac{1}{f_a}\\
g_{ai}&\equiv \dfrac{\alpha_\text{em}}{2\pi}\,\frac{c_{ai}}{v_\Phi}=\dfrac{\alpha_\text{em}}{2\pi}\,\frac{c_{ai}}{c_{agg}}\dfrac{1}{f_a}\,,
\label{gAxionEffectiveCouplings}
\end{aligned}
\ee
where $i=\{\gamma\gamma,\,ZZ,\,\gamma Z,\,WW\}$ and the traditional notation for the axion decay constant $f_a$ has been introduced.

%
%
\section{Phenomenological Features}
\label{Sect:Pheno}

Several studies have been performed to constrain the axion couplings to SM fermions and gauge bosons~\cite{Choi:1986zw,Bjorken:1988as,Carena:1988kr,Raffelt:2006cw,Adler:2008zza,Bellini:2012kz,Friedland:2012hj,Lees:2013kla,Armengaud:2013rta,Clarke:2013aya,Viaux:2013lha,Aprile:2014eoa,Ayala:2014pea,Khachatryan:2014rra,Mimasu:2014nea,Dolan:2014ska,Millea:2015qra,Vinyoles:2015aba,Aad:2015zva,Krnjaic:2015mbs,Marciano:2016yhf,Izaguirre:2016dfi,Brivio:2017ije,Bauer:2017nlg,Anastassopoulos:2017ftl,Dolan:2017osp}. Two recent summaries can be found in Refs.~\cite{Jaeckel:2015jla,Bauer:2017ris}. These bounds strongly depend on the axion mass, that also determines the decay length of the axion. The main results will be reported in this section, translating the distinct constraints into limits on the axion scale $f_a$.
\\

\noindent{\bf Coupling to photons:}\\
\indent Astrophysical, cosmological and low-energy terrestrial data provides the strongest bounds on axion couplings, those to photons (the latest constraints have been recently published in Ref.~\cite{Anastassopoulos:2017ftl}): the upper bounds on the effective couplings can be summed up as~\cite{Jaeckel:2015jla,Bauer:2017ris}
\be
\begin{aligned}
|g_{a\gamma\gamma}|\lesssim&\,7\times 10^{-11}\GeV^{-1}
&&\text{for}\,\, m_a\lesssim10 \meV
\\
|g_{a\gamma\gamma}|\lesssim&\, 10^{-10}\GeV^{-1}
&&\text{for}\,\, 10 \meV\lesssim m_a\lesssim10\eV
\\
|g_{a\gamma\gamma}|\ll&\, 10^{-12}\GeV^{-1}
&&\text{for}\,\, 10 \eV\lesssim m_a\lesssim0.1\GeV
\\
|g_{a\gamma\gamma}|\lesssim&\, 10^{-3}\GeV^{-1}
&&\text{for}\,\, 0.1 \GeV\lesssim m_a\lesssim1\TeV\,.
\end{aligned}
\label{Constraintsa2gamma}
\ee
Notice that the bounds for masses between $10\eV$ and $0.1\GeV$, which include the so-called MeV window, come from (model dependent) cosmological data~\cite{Millea:2015qra}. On the other side, for masses larger than the TeV, no constraint is present. Finally, for the mass range $0.1 \GeV\lesssim m_a\lesssim1\TeV$, the bounds may be improved by two order of magnitudes with dedicated analyses on BaBar data and at Belle-II~\cite{Mimasu:2014nea,Izaguirre:2016dfi,Dolan:2017osp}.

These bounds can be translated in terms of $f_a$ through Eq.~(\ref{gAxionEffectiveCouplings}): taking $\alpha_\text{em}=1/137.036$,
\be
\begin{aligned}
f_a\gtrsim&\,1.2\times 10^{7}\GeV
&&\text{for}\,\, m_a\lesssim10 \meV
\\
f_a\gtrsim&\,8.7\times 10^{6}\GeV
&&\text{for}\,\, 10 \meV\lesssim m_a\lesssim10\eV
\\
f_a\gg&\,8.7\times 10^{8}\GeV
&&\text{for}\,\, 10 \eV\lesssim m_a\lesssim0.1\GeV
\\
f_a\gtrsim&\, 3\GeV
&&\text{for}\,\, 0.1 \GeV\lesssim m_a\lesssim1\TeV\,.
\end{aligned}
\label{Constraintsfaa2gamma}
\ee
The first three bounds take into account the effects of the axion mixing with the $\pi^0$~\cite{diCortona:2015ldu}.\\

\noindent{\bf Coupling to gluons:}\\    
\indent Collider mono-jet searches~\cite{Khachatryan:2014rra,Mimasu:2014nea,Aad:2015zva,Brivio:2017ije} and axion-pion mixing effects~\cite{Choi:1986zw,Carena:1988kr} allows to extract bounds on the axion couplings with gluons:
\be
\begin{aligned}
|g_{agg}|\lesssim&\,1.1\times 10^{-5}\GeV^{-1}
&&\text{for}\,\,m_a\lesssim60 \MeV
\\
|g_{agg}|\lesssim&\, 10^{-4}\GeV^{-1}
&&\text{for}\,\, 60 \MeV\lesssim m_a\lesssim0.1\GeV
\end{aligned}
\ee
that can be translated into constraints on $f_a$, 
\be
\begin{aligned}
f_a\gtrsim&\,1.7\times 10^{3}\GeV
&&\text{for}\,\, m_a\lesssim60 \MeV
\\
f_a\gtrsim&\,188\GeV
&&\text{for}\,\, 60 \MeV\lesssim m_a\lesssim0.1\GeV
\end{aligned}
\ee
taking $\alpha_s(M_Z^2)=0.1184$. 
\\

\noindent{\bf Couplings to massive gauge bosons (collider):}\\   
\indent Considering LHC data with $\sqrt s=13\TeV$, dedicated analyses on Mono-$W$ ($pp\to a W(W\to \mu\nu_\mu)$) and mono-$Z$ ($pp\to aZ(Z\to ee)$) channels put bounds on axion couplings to two $W$'s and to two $Z$'s: for $0.1\lesssim m_a\lesssim1\GeV$~\cite{Brivio:2017ije}, 
\be
\begin{aligned}
|g_{aWW}|&\lesssim1.6\times10^{-3}\GeV^{-1}\\
|g_{aZZ}|&\lesssim8\times10^{-4}\GeV^{-1}\,.
\end{aligned}
\ee
A complementary analysis on LEP data~\cite{Acciarri:1994gb,Anashkin:1999da} on the radiative $Z$ decays leads to a bound on the $a\gamma Z$ coupling~\cite{Dolan:2017osp}:
\be
|g_{a\gamma Z}|\lesssim6.4\times 10^{-5}\GeV^{-1}\,.
\ee
For both the scenarios, these bounds on the effective couplings translate into the following constraints on $f_a$:
\be
\begin{aligned}
&(aWW)\qquad &&f_a\gtrsim6.4\GeV\\
&(aZZ)\qquad &&f_a\gtrsim5.7\GeV\\
&(aZ\gamma)\qquad &&f_a\gtrsim 17.8 \GeV\,.
\end{aligned}
\label{CouplingsColliderS1}
\ee

The previous bounds hold for an axion that escapes the detector and therefore is considered as missing energy in the data analysis. If instead the axion mass is sufficiently large and/or its characteristic scale $f_a$ is sufficiently low, the axion may decay within the detector and the previous limits cannot be taken into consideration. 

Considering the possibility of an axion decaying into two photons, that is typically the dominant channel,  LEP data~\cite{Acciarri:1994gb,Anashkin:1999da} on the decay $Z\to3\gamma$ has been used to constrain axion couplings with the axion decaying inside the detector. A bound on $a\gamma Z$ coupling follows from Ref.~\cite{Dolan:2017osp}: assuming that $a$ decays only into two photons,  
\be
|g_{a\gamma Z}| \lesssim6\times 10^{-4}\GeV^{-1}\,,
\ee
for axion masses in the interval $m_{\pi^0}\lesssim m_a\lesssim 10\GeV$ and
\be
|g_{a\gamma Z}| \lesssim2\times 10^{-4}\GeV^{-1}\,,
\ee
for $10\GeV\lesssim m_a\lesssim 91.2\GeV$. Considering explicitly the values for the axion couplings, the corresponding limit on $f_a$ reads
\be
\begin{aligned}
&f_a\gtrsim 1.8\GeV\quad&&\text{for}\,\, m_{\pi^0}\lesssim m_a\lesssim10\GeV\\[2mm]
&f_a\gtrsim 5.3\GeV\quad&&\text{for}\,\, 10 \GeV\lesssim m_a\lesssim91.2\GeV\,.
\end{aligned}
\ee 
A dedicated analysis with LHC data on the same observable may improve these bounds by one order of magnitude~\cite{Jaeckel:2015jla}.
\\

\noindent{\bf Couplings to fermions and $W$'s (flavour):}\\  
\indent Studies on Compton scattering of axions in the Sun, axionic recombination and de-excitation in iones and axion bremsstrahlung~\cite{Armengaud:2013rta} set very strong bounds on axion couplings to electrons for masses below $\sim80\keV$. Similar constraints are inferred from Compton conversion of solar axions~\cite{Bellini:2012kz} for masses up to $\sim10\MeV$. All together, the axion coupling to electrons is bounded to be
\be
\dfrac{c_{ae}}{c_{agg}f_a}\lesssim5.2\times10^{-8}\GeV^{-1}\,\,\text{for}\,\,1\eV\lesssim m_a\lesssim 10\MeV\,.
\ee
Even more stringent bounds arise from observation of Red Giants~\cite{Viaux:2013lha}, but for a smaller range of masses:
\be
\dfrac{c_{ae}}{c_{agg}f_a}\lesssim8.6\times10^{-10}\GeV^{-1}\,\,\text{for}\,\,m_a\lesssim 1\eV\,.
\ee
When considering the explicit value of the $c_{ae}$ coefficient, that is the same for the two PQ charge scenarios, these constraints translate into bounds on the axion scale:
\be
\begin{aligned}
f_a&\gtrsim3.9\times 10^8\GeV\,\,\text{for}\,\,m_a\lesssim 1 \eV\\
f_a&\gtrsim6.4\times 10^6\GeV\,\,\text{for}\,\,1\eV\lesssim m_a\lesssim 10\MeV\,.
\end{aligned}
\label{Constraintsfaa2ee}
\ee

Rare meson decays provide strong constraints of axion couplings to quarks and to two $W$ gauge bosons. For masses below $\sim0.2\GeV$, the most relevant observable is $K^+\to \pi^+a(a\to\text{inv.})$, whose branching ratio undergoes to the following limit~\cite{Adler:2008zza}:
\be
\mathcal{B}_{K^+\to \pi^+a(a\to\text{inv.})}<7.3\times 10^{-11}\,.
\label{BoundKdecay}
\ee
For larger masses up to a few GeVs, the \mbox{$B^+\to K^+a(a\to\text{inv.})$} decay provides the most stringent bound~\cite{Lees:2013kla}:
\be
\mathcal{B}_{B^+\to K^+a(a\to\text{inv.})}<3.2\times 10^{-5}\,.
\label{BoundBdecay}
\ee
As the axion does not couple to up-type quarks ($c_{au}=0$), the two decays $K^+\to \pi^+a$ and $B^+\to K^+a$ can only occur at 1-loop level with the axion arising from the interaction with the internal $W$ propagator. The constraints that can be deduced on $g_{aWW}$ read as~\cite{Izaguirre:2016dfi}:
\begin{align}
|g_{aWW}|\lesssim&\,3\times 10^{-6}\GeV^{-1}
&&\text{for}\,\,m_a\,\,\lesssim 0.2 \GeV
\\
|g_{aWW}|\lesssim&\, 10^{-4}\GeV^{-1}
&&\text{for}\,\, 0.2 \GeV\lesssim\,\,m_a\,\,\lesssim 5\GeV\nn
\end{align}
that can be translated in terms of $f_a$ expliciting the value of $c_{aWW}$,
\begin{align}
f_a\gtrsim&\,3.5\times10^3\GeV
&&\text{for}\,\,m_a\,\,\lesssim0.2 \GeV
\label{Constraintsfaa2WW}
\\
f_a\gtrsim&\,105\GeV
&&\text{for}\,\, 0.2 \GeV\lesssim\,\,m_a\,\,\lesssim5\GeV\,.\nn
\end{align}

Turning the attention to processes that receive 1-loop contributions with down-type quark in the internal lines, such as $D$-meson hadronic decays, no interesting bound can be extracted. The $D^+\to\pi^+a(a\to\text{inv.})$ and $D_s^+\to K^+a(a\to\text{inv.})$ decays are proportional to a combination of $c_{ad}$ and $c_{aWW}$. However, for $f_a\gtrsim105\GeV$ as identified above, the branching ratios of these processes are smaller than $10^{-12}$, that are impossible to be seen experimentally in the next future.

Finally, a recent bound from $\Upsilon\to a\gamma$ has been extracted considering bounds from BaBar and Belle~\cite{Aubert:2008as,delAmoSanchez:2010ac,Seong:2018gut}. Considering that $g_{a\gamma\gamma}\ll c_{ab}$, a bound on $c_{ad}/f_a$ can be extracted as reported in Ref.~\cite{Merlo:2019anv}:
\be
\dfrac{c_{ab}}{c_{agg}f_a}\lesssim4\times 10^{-4}\GeV^{-1}\,,
\ee
for an axion of $m_a=\mathcal{O}(1)\GeV$. This limit can be translated in terms of $f_a$ as 
\be
f_a\gtrsim830\GeV\,.
\ee
 
The previous bounds are valid only for a stable axion at detector size. If instead the axion further decays, present data from $b\to s g$ or $b\to sq\ov{q}$ from CLEO collaboration~\cite{Coan:1997ye} allows to put a bound on axion couplings to b quarks:
\be
\dfrac{c_{ad}}{c_{agg}f_a}\lesssim5\times 10^{-4}\GeV^{-1}\,,
\ee
for $0.4\lesssim m_a\lesssim4.8\GeV$. This constraint translates into a bound on $f_a$ that reads
\be
f_a\gtrsim667\GeV\,.
\ee

Similar bounds can be inferred with a future analysis on $B^\pm\to K^\pm\, a(\to2\gamma)$ at Belle-II~\cite{Izaguirre:2016dfi}. In case the branching ratio for this observables will be measured at the level of $10^{-6}$, then values of $f_a$ as large as $550\GeV$ in the $aWW$ coupling could be tested.\\

\noindent{\bf The axion mass and the ALP scenario:}\\  
\indent Without an explicit soft breaking source of the shift symmetry, a mass term for the MFV axion may arise, as for the traditional QCD axion, from non-perturbative dynamics: the axion mixing with neutral mesons induces a contribution which is estimated to be~\cite{Shifman:1979if,Bardeen:1978nq,DiVecchia:1980yfw}
\be
m_a\sim 6 \mueV\left(\dfrac{10^{12}\GeV}{f_a}\right)\,,
\label{AxionMassContributionQCD}
\ee
and not much larger than a few eV. Additional contributions may arise {\it \`a la} KSVZ axion in the presence of exotic fermions that couple to the axion. Exotic fermions are typically present when constructing the underlying theory originating the effective terms in Eq.~(\ref{YukawaAxionLagrangian}) (see for example Ref.~\cite{Varzielas:2010mp}) or are required from anomaly cancellation conditions in models with gauged flavour symmetries~\cite{Grinstein:2010ve,Feldmann:2010yp,Guadagnoli:2011id,Buras:2011zb,Buras:2011wi,Feldmann:2016hvo,Alonso:2016onw}: the largest mass contribution originated in these cases is of hundreds of eV, for values of the axion scale $f_a$ close to the TeV. Even considering possible contributions of this kind, one can safely conclude that the MFV axion mass is smaller than the keV, unless explicit shift symmetry breaking sources are introduced in the scalar potential. For these mass values the strongest constraints arise from the axion coupling to photons, Eq.~(\ref{Constraintsfaa2gamma}), and to electrons, Eq.~(\ref{Constraintsfaa2ee}): the axion scale is necessarily larger than $\sim10^{10}\GeV$ and $\sim10^{9}\GeV$, preventing any possibility to detect the MFV axion at colliders or in flavour searches. 

On the other side, if a signal of detection that may be interpreted in terms of an axion is seen, it may be compatible with the MFV axion at the price of invoking an explicit breaking of the shift symmetry (gravitational and/or Planck-scale effects \cite{Barr:1992qq,Kamionkowski:1992mf,Holman:1992us,Alonso:2017avz} are examples of unavoidable explicit breaking sources, but the corresponding mass contributions are tiny): in this case, the relation between the axion mass and its scale gets broken and the bounds aforementioned may be avoided. In the common language, this eventuality is refereed to as Axion-like-particle (ALP) framework, that received much attention from the community in the last years. 

In what follows, this last scenario will be considered, assuming a MFV axion mass much larger than the eV region. For masses of the order of the GeV, the strong bounds from the $a\gamma\gamma$ and $aee$ couplings are easily evaded, and the next most sensitive observables are those from collider and from flavour. For even larger masses, no bounds at all have been put. 

By increasing the axion mass and/or lowering the scale $f_a$, however, its decay length decreases, and the axion may decay within the detector: in this case, some of the previous listed bounds cannot apply anymore. The distance travelled by an axion after being produced can be casted in the following expression~\cite{Brivio:2017ije}:
\be
d\approx\dfrac{10^4}{c^2_{ai}}\left(\dfrac{\MeV}{m_a}\right)^4\left(\dfrac{f_a}{\GeV}\right)^2\left(\dfrac{|p_a|}{\GeV}\right)\m\,,
\ee
where the typical momentum considered is of $100\GeV$. Selecting a benchmark region with $m_a\simeq1\GeV$ and $f_a=1\TeV$, the axion may decay into two photons, two gluons, or two light fermions. Once considering the values for $c_{ai}$ as reported in Tab.~\ref{tab:pheno}, the dominant channel is the radiative one (see i.e. Ref.~\cite{Bauer:2017ris} for the relevant expressions of the axion decays). The decay length for this benchmark axion turns out to be slightly larger than $1\mm$. The most sensitive observables to the this ALP is $b\to s g$ from CLEO collaboration and $Z\to3\gamma$ from LEP and LHC experiments: indeed, these processes are sensitive to values of $f_a$ up to $\sim1\TeV$.\\

\noindent{\bf Comparison with the Axiflavon:}\\  
\indent The Axiflavon~\cite{Ema:2016ops,Calibbi:2016hwq} is the axion arising in the context of the FN mechanism and has flavour violating couplings, in the mass basis for fermions, predicted in terms of the FN charges, up to $\cO(1)$ uncertainties.  This represents a major difference with respect to the MFV axion: the presence of flavour violating couplings induces tree-level flavour changing neutral current processes, such as the meson decays described in the previous section. To satisfy the present bounds on $K$ and $B$ decays, the axion scale $f_a$ needs to be of the order of $10^{10}\GeV$~\cite{Calibbi:2016hwq}, that approximatively coincides with the values necessary to pass the very stringent bounds on the $a\gamma\gamma$ and $aee$ couplings. The Axiflavon is therefore an example of visible QCD axion, as it predicts low-energy flavour effects, despite of the very large value of the axion scale $f_a$. On the other side, no signals are expected at colliders, as indeed effects in mono-$W$ and mono-$Z$ channels, and in the $Z$ boson width are expected to be tiny and not appreciable nor in the future phases of LHC neither in next generation of linear/circular colliders. As a final comment that helps distinguishing between the MFV axion and the Axiflavon is the prediction for the ratio between the axion coupling to photons and that to gluons: in the first model this ratio is strictly predicted to be $8/3$, while in the second one it may vary within the range $[2.4,\,3]$.

%
%
\section{Conclusions}
\label{Sect:Conclusions}

The MFV ansatz, beside leading to very predictive context to solve the BSM flavour problem, is a fascinating approach to attempt to explain the flavour puzzle. Besides the non-Abelian parts of the full flavour symmetry, the Abelian terms may be responsible for explaining the mass hierarchies between the third fermion families. The fermion charge assignment is however not vectorial and this opens the possibility to interpret the Goldstone boson arising from the spontaneous breaking of one of these terms as an axion that solves the Strong CP problem. 

The MFV axion couplings are determined by the fermion charge assignments, which are almost all fixed by requiring that $m_t\sim v/\sqrt2$, $m_b/m_t$ and $m_\tau/m_t$ fit the measured values, and the predicted value for $\mu\to e$ conversion in golden nuclei does not saturate the present experimental bound. The axion couplings with up-type quarks are identically vanishing, while those with down-type quarks, charged leptons, two gluons, two photons, two $Z$'s, $Z\gamma$ are all non-vanishing and fixed to specific values. The only freedom left is in the value of the axion coupling to two $W$'s: the choice for the charge of the lepton doublet $x_\ell$ is not unique, but it can take the value $x_\ell=0$, that identified the scenario $S0$ where $aWW$ couplings is also vanishing, or the value $x_\ell=1$, dubbed as $S1$ scenario where the MFV axion does couple to two $W$'s.

The most constraining bounds affect the axion couplings to photons and to electrons: all in all, the axion scale $f_a$ needs to be larger than $10^8\GeV$ for masses smaller than $0.1\GeV$. If follows that a MFV axion with masses below this value can be considered an invisible axion, as no effects are expected neither in low-energy experiments nor at colliders. With respect to other invisible axion models, such as DFSZ or KSVZ, the MFV axion has the advantages that its origin is not linked to an {\it ad hoc} introduction of the PQ symmetry, but follows naturally from an Abelian term of the SM flavour symmetry.

If a signal is seen in present or near future experiments that may be interpreted in terms of an axion, it would still be compatible with the MFV axion, but at the price of breaking the proportionality between its mass and its characteristic scale, invoking an explicit shift symmetry breaking. Indeed, for $m_a\simeq1\GeV$, the bounds from $a\gamma\gamma$ and $aee$ couplings do not apply, opening the possibility of low-energy signals. For a stable axion, the most sensitive observables are those from flavour: $B\to K a(a\to\text{inv.})$ decay ($K\to\pi a(a\to\text{inv.})$ one is forbidden kinematically for this benchmark) and $\Upsilon\to a\gamma$ .  For an axion decaying into the detector, the dominant channel being in two photons, LEP data on $Z\to3\gamma$ is currently putting the strongest bounds, but that will be improved by one order of magnitude with a dedicated LHC analysis on the same observable and with a Belle-II analysis on $B^\pm\to K^\pm\, a(\to2\gamma)$ decay. Considering the axion-bottom quark coupling, CLEO data on $b\to sg$ is currently putting the strongest bound on a decaying axion.

In the MFV axion model, the axion couplings are flavour conserving (but flavour universality violating): in consequence, this axion does not give rise to flavour changing neutral current processes at tree level, but describes flavour changing processes at 1-loop level. This represents the major difference with respect to the Axiflavon model, where the axion does violate flavour and describes rare meson decays at tree level: the existing bounds from  $K\to\pi a$ and $B\to K a$ decays constrain the Axiflavon scale to be larger than $10^{10}\GeV$: for these values the bounds on $a\gamma\gamma$ and $aee$ are satisfied, while all the effects at colliders are expected to be tiny and far from the expected future improvements. 

To summarise, if no signal at all will appear neither at colliders nor in low-energy flavour experiments, then the only possibility is the one of an invisible axion with a very large scale $f_a$, being the DFSZ, the KSVZ, the MFV axion and the Axiflavon all equally viable.

If a signal emerges at colliders and flavour factories, then this would be in favour of a heavy MFV axion, while disfavouring the Axiflavon. If, instead, a signal emerges only at flavour factories, it will be the other way around. A precise measure of the axion couplings to photons and to gluons may be a smoking gun for the MFV axion model as the ratio between these two couplings is strictly predicted to be $8/3$. 

\acknowledgments
We thank Enrique Fern\'andez Mart\'inez and Emmanuel Stamou for enjoyable brainstorming during the development of this project and Paride Paradisi and Carlos Pena Ruano for valuable conversations on meson decays and flavoured axion. L.M. thanks the department of Physics and Astronomy of the Universit\`a degli Studi di Padova and the Fermilab Theory Division for hospitality during the writing up of this paper. L.M. acknowledges partial financial support by the European Union's Horizon 2020 research and innovation programme under the Marie Sklodowska-Curie grant agreements No 690575 and No 674896, by the Spanish MINECO through the ``Ram\'on y Cajal'' programme (RYC-2015-17173), and by the Spanish ``Agencia Estatal de Investigaci\'on'' (AEI) and the EU ``Fondo Europeo de Desarrollo Regional'' (FEDER) through the project FPA2016-78645-P, and through the Centro de excelencia Severo Ochoa Program under grant SEV-2016-0597.

%
%

\begin{thebibliography}{100}

\bibitem{Froggatt:1978nt}
C.~D. Froggatt and H.~B. Nielsen,  Nucl. Phys. {\bf B147} (1979) 277--298.

\bibitem{Altarelli:2000fu}
G.~Altarelli, F.~Feruglio, and I.~Masina,  JHEP {\bf 11} (2000) 040,
  [\href{http://xxx.lanl.gov/abs/hep-ph/0007254}{{\tt hep-ph/0007254}}].

\bibitem{Altarelli:2002sg}
G.~Altarelli, F.~Feruglio, and I.~Masina,  JHEP {\bf 01} (2003) 035,
  [\href{http://xxx.lanl.gov/abs/hep-ph/0210342}{{\tt hep-ph/0210342}}].

\bibitem{Chankowski:2005qp}
P.~H. Chankowski, K.~Kowalska, S.~Lavignac, and S.~Pokorski,  Phys. Rev. {\bf
  D71} (2005) 055004, [\href{http://xxx.lanl.gov/abs/hep-ph/0501071}{{\tt
  hep-ph/0501071}}].

\bibitem{Buchmuller:2011tm}
W.~Buchmuller, V.~Domcke, and K.~Schmitz,  JHEP {\bf 03} (2012) 008,
  [\href{http://xxx.lanl.gov/abs/1111.3872}{{\tt arXiv:1111.3872}}].

\bibitem{Altarelli:2012ia}
G.~Altarelli, F.~Feruglio, I.~Masina, and L.~Merlo,  JHEP {\bf 11} (2012) 139,
  [\href{http://xxx.lanl.gov/abs/1207.0587}{{\tt arXiv:1207.0587}}].

\bibitem{Bergstrom:2014owa}
J.~Bergstrom, D.~Meloni, and L.~Merlo,  Phys. Rev. {\bf D89} (2014), no.~9
  093021, [\href{http://xxx.lanl.gov/abs/1403.4528}{{\tt arXiv:1403.4528}}].

\bibitem{Ma:2001dn}
E.~Ma and G.~Rajasekaran,  Phys. Rev. {\bf D64} (2001) 113012,
  [\href{http://xxx.lanl.gov/abs/hep-ph/0106291}{{\tt hep-ph/0106291}}].

\bibitem{Babu:2002dz}
K.~S. Babu, E.~Ma, and J.~W.~F. Valle,  Phys. Lett. {\bf B552} (2003) 207--213,
  [\href{http://xxx.lanl.gov/abs/hep-ph/0206292}{{\tt hep-ph/0206292}}].

\bibitem{Altarelli:2005yp}
G.~Altarelli and F.~Feruglio,  Nucl. Phys. {\bf B720} (2005) 64--88,
  [\href{http://xxx.lanl.gov/abs/hep-ph/0504165}{{\tt hep-ph/0504165}}].

\bibitem{Altarelli:2005yx}
G.~Altarelli and F.~Feruglio,  Nucl. Phys. {\bf B741} (2006) 215--235,
  [\href{http://xxx.lanl.gov/abs/hep-ph/0512103}{{\tt hep-ph/0512103}}].

\bibitem{Feruglio:2007uu}
F.~Feruglio, C.~Hagedorn, Y.~Lin, and L.~Merlo,  Nucl. Phys. {\bf B775} (2007)
  120--142, [\href{http://xxx.lanl.gov/abs/hep-ph/0702194}{{\tt
  hep-ph/0702194}}]. [Erratum: Nucl. Phys.B836,127(2010)].

\bibitem{Bazzocchi:2009pv}
F.~Bazzocchi, L.~Merlo, and S.~Morisi,  Nucl. Phys. {\bf B816} (2009) 204--226,
  [\href{http://xxx.lanl.gov/abs/0901.2086}{{\tt arXiv:0901.2086}}].

\bibitem{Bazzocchi:2009da}
F.~Bazzocchi, L.~Merlo, and S.~Morisi,  Phys. Rev. {\bf D80} (2009) 053003,
  [\href{http://xxx.lanl.gov/abs/0902.2849}{{\tt arXiv:0902.2849}}].

\bibitem{Altarelli:2009gn}
G.~Altarelli, F.~Feruglio, and L.~Merlo,  JHEP {\bf 05} (2009) 020,
  [\href{http://xxx.lanl.gov/abs/0903.1940}{{\tt arXiv:0903.1940}}].

\bibitem{Toorop:2010yh}
R.~de~Adelhart~Toorop, F.~Bazzocchi, and L.~Merlo,  JHEP {\bf 08} (2010) 001,
  [\href{http://xxx.lanl.gov/abs/1003.4502}{{\tt arXiv:1003.4502}}].

\bibitem{Altarelli:2010gt}
G.~Altarelli and F.~Feruglio,  Rev. Mod. Phys. {\bf 82} (2010) 2701--2729,
  [\href{http://xxx.lanl.gov/abs/1002.0211}{{\tt arXiv:1002.0211}}].

\bibitem{Varzielas:2010mp}
I.~de~Medeiros~Varzielas and L.~Merlo,  JHEP {\bf 02} (2011) 062,
  [\href{http://xxx.lanl.gov/abs/1011.6662}{{\tt arXiv:1011.6662}}].

\bibitem{Toorop:2011jn}
R.~de~Adelhart~Toorop, F.~Feruglio, and C.~Hagedorn,  Phys. Lett. {\bf B703}
  (2011) 447--451, [\href{http://xxx.lanl.gov/abs/1107.3486}{{\tt
  arXiv:1107.3486}}].

\bibitem{Grimus:2011fk}
W.~Grimus and P.~O. Ludl,  J. Phys. {\bf A45} (2012) 233001,
  [\href{http://xxx.lanl.gov/abs/1110.6376}{{\tt arXiv:1110.6376}}].

\bibitem{deAdelhartToorop:2011re}
R.~de~Adelhart~Toorop, F.~Feruglio, and C.~Hagedorn,  Nucl. Phys. {\bf B858}
  (2012) 437--467, [\href{http://xxx.lanl.gov/abs/1112.1340}{{\tt
  arXiv:1112.1340}}].

\bibitem{King:2011ab}
S.~F. King and C.~Luhn,  JHEP {\bf 03} (2012) 036,
  [\href{http://xxx.lanl.gov/abs/1112.1959}{{\tt arXiv:1112.1959}}].

\bibitem{Altarelli:2012ss}
G.~Altarelli, F.~Feruglio, and L.~Merlo,  Fortsch. Phys. {\bf 61} (2013)
  507--534, [\href{http://xxx.lanl.gov/abs/1205.5133}{{\tt arXiv:1205.5133}}].

\bibitem{Bazzocchi:2012st}
F.~Bazzocchi and L.~Merlo,  Fortsch. Phys. {\bf 61} (2013) 571--596,
  [\href{http://xxx.lanl.gov/abs/1205.5135}{{\tt arXiv:1205.5135}}].

\bibitem{King:2013eh}
S.~F. King and C.~Luhn,  Rept. Prog. Phys. {\bf 76} (2013) 056201,
  [\href{http://xxx.lanl.gov/abs/1301.1340}{{\tt arXiv:1301.1340}}].

\bibitem{Feruglio:2008ht}
F.~Feruglio, C.~Hagedorn, Y.~Lin, and L.~Merlo,  Nucl. Phys. {\bf B809} (2009)
  218--243, [\href{http://xxx.lanl.gov/abs/0807.3160}{{\tt arXiv:0807.3160}}].

\bibitem{Feruglio:2009iu}
F.~Feruglio, C.~Hagedorn, and L.~Merlo,  JHEP {\bf 03} (2010) 084,
  [\href{http://xxx.lanl.gov/abs/0910.4058}{{\tt arXiv:0910.4058}}].

\bibitem{Lin:2009sq}
Y.~Lin, L.~Merlo, and A.~Paris,  Nucl. Phys. {\bf B835} (2010) 238--261,
  [\href{http://xxx.lanl.gov/abs/0911.3037}{{\tt arXiv:0911.3037}}].

\bibitem{Feruglio:2009hu}
F.~Feruglio, C.~Hagedorn, Y.~Lin, and L.~Merlo,  Nucl. Phys. {\bf B832} (2010)
  251--288, [\href{http://xxx.lanl.gov/abs/0911.3874}{{\tt arXiv:0911.3874}}].

\bibitem{Ishimori:2010au}
H.~Ishimori, T.~Kobayashi, H.~Ohki, Y.~Shimizu, H.~Okada, and M.~Tanimoto,
  Prog. Theor. Phys. Suppl. {\bf 183} (2010) 1--163,
  [\href{http://xxx.lanl.gov/abs/1003.3552}{{\tt arXiv:1003.3552}}].

\bibitem{Toorop:2010ex}
R.~de~Adelhart~Toorop, F.~Bazzocchi, L.~Merlo, and A.~Paris,  JHEP {\bf 03}
  (2011) 035, [\href{http://xxx.lanl.gov/abs/1012.1791}{{\tt
  arXiv:1012.1791}}]. [Erratum: JHEP01,098(2013)].

\bibitem{Toorop:2010kt}
R.~de~Adelhart~Toorop, F.~Bazzocchi, L.~Merlo, and A.~Paris,  JHEP {\bf 03}
  (2011) 040, [\href{http://xxx.lanl.gov/abs/1012.2091}{{\tt
  arXiv:1012.2091}}].

\bibitem{Merlo:2011hw}
L.~Merlo, S.~Rigolin, and B.~Zaldivar,  JHEP {\bf 11} (2011) 047,
  [\href{http://xxx.lanl.gov/abs/1108.1795}{{\tt arXiv:1108.1795}}].

\bibitem{Altarelli:2012bn}
G.~Altarelli, F.~Feruglio, L.~Merlo, and E.~Stamou,  JHEP {\bf 08} (2012) 021,
  [\href{http://xxx.lanl.gov/abs/1205.4670}{{\tt arXiv:1205.4670}}].

\bibitem{Abe:2011sj}
{\bf T2K} Collaboration, K.~Abe {\em et.~al.},  Phys. Rev. Lett. {\bf 107}
  (2011) 041801, [\href{http://xxx.lanl.gov/abs/1106.2822}{{\tt
  arXiv:1106.2822}}].

\bibitem{Adamson:2011qu}
{\bf MINOS} Collaboration, P.~Adamson {\em et.~al.},  Phys. Rev. Lett. {\bf
  107} (2011) 181802, [\href{http://xxx.lanl.gov/abs/1108.0015}{{\tt
  arXiv:1108.0015}}].

\bibitem{Abe:2011fz}
{\bf Double Chooz} Collaboration, Y.~Abe {\em et.~al.},  Phys. Rev. Lett. {\bf
  108} (2012) 131801, [\href{http://xxx.lanl.gov/abs/1112.6353}{{\tt
  arXiv:1112.6353}}].

\bibitem{An:2012eh}
{\bf Daya Bay} Collaboration, F.~P. An {\em et.~al.},  Phys. Rev. Lett. {\bf
  108} (2012) 171803, [\href{http://xxx.lanl.gov/abs/1203.1669}{{\tt
  arXiv:1203.1669}}].

\bibitem{Ahn:2012nd}
{\bf RENO} Collaboration, J.~K. Ahn {\em et.~al.},  Phys. Rev. Lett. {\bf 108}
  (2012) 191802, [\href{http://xxx.lanl.gov/abs/1204.0626}{{\tt
  arXiv:1204.0626}}].

\bibitem{Chivukula:1987py}
R.~S. Chivukula and H.~Georgi,  Phys. Lett. {\bf B188} (1987) 99--104.

\bibitem{DAmbrosio:2002vsn}
G.~D'Ambrosio, G.~F. Giudice, G.~Isidori, and A.~Strumia,  Nucl. Phys. {\bf
  B645} (2002) 155--187, [\href{http://xxx.lanl.gov/abs/hep-ph/0207036}{{\tt
  hep-ph/0207036}}].

\bibitem{Cirigliano:2005ck}
V.~Cirigliano, B.~Grinstein, G.~Isidori, and M.~B. Wise,  Nucl. Phys. {\bf
  B728} (2005) 121--134, [\href{http://xxx.lanl.gov/abs/hep-ph/0507001}{{\tt
  hep-ph/0507001}}].

\bibitem{Davidson:2006bd}
S.~Davidson and F.~Palorini,  Phys. Lett. {\bf B642} (2006) 72--80,
  [\href{http://xxx.lanl.gov/abs/hep-ph/0607329}{{\tt hep-ph/0607329}}].

\bibitem{Alonso:2011jd}
R.~Alonso, G.~Isidori, L.~Merlo, L.~A. Munoz, and E.~Nardi,  JHEP {\bf 06}
  (2011) 037, [\href{http://xxx.lanl.gov/abs/1103.5461}{{\tt
  arXiv:1103.5461}}].

\bibitem{Dinh:2017smk}
D.~N. Dinh, L.~Merlo, S.~T. Petcov, and R.~Vega-Álvarez,  JHEP {\bf 07} (2017)
  089, [\href{http://xxx.lanl.gov/abs/1705.09284}{{\tt arXiv:1705.09284}}].

\bibitem{Cirigliano:2006su}
V.~Cirigliano and B.~Grinstein,  Nucl. Phys. {\bf B752} (2006) 18--39,
  [\href{http://xxx.lanl.gov/abs/hep-ph/0601111}{{\tt hep-ph/0601111}}].

\bibitem{Grinstein:2006cg}
B.~Grinstein, V.~Cirigliano, G.~Isidori, and M.~B. Wise,  Nucl. Phys. {\bf
  B763} (2007) 35--48, [\href{http://xxx.lanl.gov/abs/hep-ph/0608123}{{\tt
  hep-ph/0608123}}].

\bibitem{Hurth:2008jc}
T.~Hurth, G.~Isidori, J.~F. Kamenik, and F.~Mescia,  Nucl. Phys. {\bf B808}
  (2009) 326--346, [\href{http://xxx.lanl.gov/abs/0807.5039}{{\tt
  arXiv:0807.5039}}].

\bibitem{Kagan:2009bn}
A.~L. Kagan, G.~Perez, T.~Volansky, and J.~Zupan,  Phys. Rev. {\bf D80} (2009)
  076002, [\href{http://xxx.lanl.gov/abs/0903.1794}{{\tt arXiv:0903.1794}}].

\bibitem{Gavela:2009cd}
M.~B. Gavela, T.~Hambye, D.~Hernandez, and P.~Hernandez,  JHEP {\bf 09} (2009)
  038, [\href{http://xxx.lanl.gov/abs/0906.1461}{{\tt arXiv:0906.1461}}].

\bibitem{Grinstein:2010ve}
B.~Grinstein, M.~Redi, and G.~Villadoro,  JHEP {\bf 11} (2010) 067,
  [\href{http://xxx.lanl.gov/abs/1009.2049}{{\tt arXiv:1009.2049}}].

\bibitem{Feldmann:2010yp}
T.~Feldmann,  JHEP {\bf 04} (2011) 043,
  [\href{http://xxx.lanl.gov/abs/1010.2116}{{\tt arXiv:1010.2116}}].

\bibitem{Guadagnoli:2011id}
D.~Guadagnoli, R.~N. Mohapatra, and I.~Sung,  JHEP {\bf 04} (2011) 093,
  [\href{http://xxx.lanl.gov/abs/1103.4170}{{\tt arXiv:1103.4170}}].

\bibitem{Buras:2011zb}
A.~J. Buras, L.~Merlo, and E.~Stamou,  JHEP {\bf 08} (2011) 124,
  [\href{http://xxx.lanl.gov/abs/1105.5146}{{\tt arXiv:1105.5146}}].

\bibitem{Buras:2011wi}
A.~J. Buras, M.~V. Carlucci, L.~Merlo, and E.~Stamou,  JHEP {\bf 03} (2012)
  088, [\href{http://xxx.lanl.gov/abs/1112.4477}{{\tt arXiv:1112.4477}}].

\bibitem{Alonso:2012jc}
R.~Alonso, M.~B. Gavela, L.~Merlo, S.~Rigolin, and J.~Yepes,  JHEP {\bf 06}
  (2012) 076, [\href{http://xxx.lanl.gov/abs/1201.1511}{{\tt
  arXiv:1201.1511}}].

\bibitem{Isidori:2012ts}
G.~Isidori and D.~M. Straub,  Eur. Phys. J. {\bf C72} (2012) 2103,
  [\href{http://xxx.lanl.gov/abs/1202.0464}{{\tt arXiv:1202.0464}}].

\bibitem{Lopez-Honorez:2013wla}
L.~Lopez-Honorez and L.~Merlo,  Phys. Lett. {\bf B722} (2013) 135--143,
  [\href{http://xxx.lanl.gov/abs/1303.1087}{{\tt arXiv:1303.1087}}].

\bibitem{Bishara:2015mha}
F.~Bishara, A.~Greljo, J.~F. Kamenik, E.~Stamou, and J.~Zupan,  JHEP {\bf 12}
  (2015) 130, [\href{http://xxx.lanl.gov/abs/1505.03862}{{\tt
  arXiv:1505.03862}}].

\bibitem{Lee:2015qra}
C.-J. Lee and J.~Tandean,  JHEP {\bf 08} (2015) 123,
  [\href{http://xxx.lanl.gov/abs/1505.04692}{{\tt arXiv:1505.04692}}].

\bibitem{Feldmann:2016hvo}
T.~Feldmann, C.~Luhn, and P.~Moch,  JHEP {\bf 11} (2016) 078,
  [\href{http://xxx.lanl.gov/abs/1608.04124}{{\tt arXiv:1608.04124}}].

\bibitem{Alonso:2016onw}
R.~Alonso, E.~Fernandez~Mart{\'\i ne}z, M.~B. Gavela, B.~Grinstein, L.~Merlo,
  and P.~Quilez,  JHEP {\bf 12} (2016) 119,
  [\href{http://xxx.lanl.gov/abs/1609.05902}{{\tt arXiv:1609.05902}}].

\bibitem{Merlo:2018rin}
L.~Merlo and S.~Rosauro-Alcaraz,  JHEP {\bf 07} (2018) 036,
  [\href{http://xxx.lanl.gov/abs/1801.03937}{{\tt arXiv:1801.03937}}].

\bibitem{Isidori:2010kg}
G.~Isidori, Y.~Nir, and G.~Perez,  Ann. Rev. Nucl. Part. Sci. {\bf 60} (2010)
  355, [\href{http://xxx.lanl.gov/abs/1002.0900}{{\tt arXiv:1002.0900}}].

\bibitem{Alonso:2011yg}
R.~Alonso, M.~B. Gavela, L.~Merlo, and S.~Rigolin,  JHEP {\bf 07} (2011) 012,
  [\href{http://xxx.lanl.gov/abs/1103.2915}{{\tt arXiv:1103.2915}}].

\bibitem{Alonso:2012fy}
R.~Alonso, M.~B. Gavela, D.~Hernandez, and L.~Merlo,  Phys. Lett. {\bf B715}
  (2012) 194--198, [\href{http://xxx.lanl.gov/abs/1206.3167}{{\tt
  arXiv:1206.3167}}].

\bibitem{Alonso:2013mca}
R.~Alonso, M.~B. Gavela, D.~Hernández, L.~Merlo, and S.~Rigolin,  JHEP {\bf
  08} (2013) 069, [\href{http://xxx.lanl.gov/abs/1306.5922}{{\tt
  arXiv:1306.5922}}].

\bibitem{Alonso:2013nca}
R.~Alonso, M.~B. Gavela, G.~Isidori, and L.~Maiani,  JHEP {\bf 11} (2013) 187,
  [\href{http://xxx.lanl.gov/abs/1306.5927}{{\tt arXiv:1306.5927}}].

\bibitem{Anselm:1996jm}
A.~Anselm and Z.~Berezhiani,  Nucl. Phys. {\bf B484} (1997) 97--123,
  [\href{http://xxx.lanl.gov/abs/hep-ph/9605400}{{\tt hep-ph/9605400}}].

\bibitem{Barbieri:1999km}
R.~Barbieri, L.~J. Hall, G.~L. Kane, and G.~G. Ross,
  \href{http://xxx.lanl.gov/abs/hep-ph/9901228}{{\tt hep-ph/9901228}}.

\bibitem{Berezhiani:2001mh}
Z.~Berezhiani and A.~Rossi,  Nucl. Phys. Proc. Suppl. {\bf 101} (2001)
  410--420, [\href{http://xxx.lanl.gov/abs/hep-ph/0107054}{{\tt
  hep-ph/0107054}}]. [,410(2001)].

\bibitem{Feldmann:2009dc}
T.~Feldmann, M.~Jung, and T.~Mannel,  Phys. Rev. {\bf D80} (2009) 033003,
  [\href{http://xxx.lanl.gov/abs/0906.1523}{{\tt arXiv:0906.1523}}].

\bibitem{Nardi:2011st}
E.~Nardi,  Phys. Rev. {\bf D84} (2011) 036008,
  [\href{http://xxx.lanl.gov/abs/1105.1770}{{\tt arXiv:1105.1770}}].

\bibitem{Peccei:1977hh}
R.~D. Peccei and H.~R. Quinn,  Phys. Rev. Lett. {\bf 38} (1977) 1440--1443.

\bibitem{Wilczek:1977pj}
F.~Wilczek,  Phys. Rev. Lett. {\bf 40} (1978) 279--282.

\bibitem{Weinberg:1977ma}
S.~Weinberg,  Phys. Rev. Lett. {\bf 40} (1978) 223--226.

\bibitem{diCortona:2015ldu}
G.~Grilli~di Cortona, E.~Hardy, J.~Pardo~Vega, and G.~Villadoro,  JHEP {\bf 01}
  (2016) 034, [\href{http://xxx.lanl.gov/abs/1511.02867}{{\tt
  arXiv:1511.02867}}].

\bibitem{Kim:1979if}
J.~E. Kim,  Phys. Rev. Lett. {\bf 43} (1979) 103.

\bibitem{Shifman:1979if}
M.~A. Shifman, A.~I. Vainshtein, and V.~I. Zakharov,  Nucl. Phys. {\bf B166}
  (1980) 493--506.

\bibitem{Zhitnitsky:1980tq}
A.~R. Zhitnitsky,  Sov. J. Nucl. Phys. {\bf 31} (1980) 260. [Yad.
  Fiz.31,497(1980)].

\bibitem{Dine:1981rt}
M.~Dine, W.~FisCHLer, and M.~Srednicki,  Phys. Lett. {\bf B104} (1981)
  199--202.

\bibitem{Ema:2016ops}
Y.~Ema, K.~Hamaguchi, T.~Moroi, and K.~Nakayama,  JHEP {\bf 01} (2017) 096,
  [\href{http://xxx.lanl.gov/abs/1612.05492}{{\tt arXiv:1612.05492}}].

\bibitem{Calibbi:2016hwq}
L.~Calibbi, F.~Goertz, D.~Redigolo, R.~Ziegler, and J.~Zupan,  Phys. Rev. {\bf
  D95} (2017), no.~9 095009, [\href{http://xxx.lanl.gov/abs/1612.08040}{{\tt
  arXiv:1612.08040}}].

\bibitem{Wilczek:1982rv}
F.~Wilczek,  Phys. Rev. Lett. {\bf 49} (1982) 1549--1552.

\bibitem{Albrecht:2010xh}
M.~E. Albrecht, T.~Feldmann, and T.~Mannel,  JHEP {\bf 10} (2010) 089,
  [\href{http://xxx.lanl.gov/abs/1002.4798}{{\tt arXiv:1002.4798}}].

\bibitem{Weinberg:1979sa}
S.~Weinberg,  Phys. Rev. Lett. {\bf 43} (1979) 1566--1570.

\bibitem{Choi:2017gpf}
K.~Choi, S.~H. Im, C.~B. Park, and S.~Yun,
  \href{http://xxx.lanl.gov/abs/1708.00021}{{\tt arXiv:1708.00021}}.

\bibitem{Kaplan:1983fs}
D.~B. Kaplan and H.~Georgi,  Phys. Lett. {\bf B136} (1984) 183--186.

\bibitem{Kaplan:1983sm}
D.~B. Kaplan, H.~Georgi, and S.~Dimopoulos,  Phys. Lett. {\bf B136} (1984)
  187--190.

\bibitem{Banks:1984gj}
T.~Banks,  Nucl. Phys. {\bf B243} (1984) 125--130.

\bibitem{Agashe:2004rs}
K.~Agashe, R.~Contino, and A.~Pomarol,  Nucl. Phys. {\bf B719} (2005) 165--187,
  [\href{http://xxx.lanl.gov/abs/hep-ph/0412089}{{\tt hep-ph/0412089}}].

\bibitem{Gripaios:2009pe}
B.~Gripaios, A.~Pomarol, F.~Riva, and J.~Serra,  JHEP {\bf 04} (2009) 070,
  [\href{http://xxx.lanl.gov/abs/0902.1483}{{\tt arXiv:0902.1483}}].

\bibitem{Mrazek:2011iu}
J.~Mrazek, A.~Pomarol, R.~Rattazzi, M.~Redi, J.~Serra, and A.~Wulzer,  Nucl.
  Phys. {\bf B853} (2011) 1--48, [\href{http://xxx.lanl.gov/abs/1105.5403}{{\tt
  arXiv:1105.5403}}].

\bibitem{Alonso:2014wta}
R.~Alonso, I.~Brivio, B.~Gavela, L.~Merlo, and S.~Rigolin,  JHEP {\bf 12}
  (2014) 034, [\href{http://xxx.lanl.gov/abs/1409.1589}{{\tt
  arXiv:1409.1589}}].

\bibitem{Panico:2015jxa}
G.~Panico and A.~Wulzer,  Lect. Notes Phys. {\bf 913} (2016) pp.1--316,
  [\href{http://xxx.lanl.gov/abs/1506.01961}{{\tt arXiv:1506.01961}}].

\bibitem{Hierro:2015nna}
I.~M. Hierro, L.~Merlo, and S.~Rigolin,  JHEP {\bf 04} (2016) 016,
  [\href{http://xxx.lanl.gov/abs/1510.07899}{{\tt arXiv:1510.07899}}].

\bibitem{Halyo:1991pc}
E.~Halyo,  Mod. Phys. Lett. {\bf A8} (1993) 275--284.

\bibitem{Goldberger:2008zz}
W.~D. Goldberger, B.~Grinstein, and W.~Skiba,  Phys. Rev. Lett. {\bf 100}
  (2008) 111802, [\href{http://xxx.lanl.gov/abs/0708.1463}{{\tt
  arXiv:0708.1463}}].

\bibitem{Vecchi:2010gj}
L.~Vecchi,  Phys. Rev. {\bf D82} (2010) 076009,
  [\href{http://xxx.lanl.gov/abs/1002.1721}{{\tt arXiv:1002.1721}}].

\bibitem{Matsuzaki:2012mk}
S.~Matsuzaki and K.~Yamawaki,  Phys. Lett. {\bf B719} (2013) 378--382,
  [\href{http://xxx.lanl.gov/abs/1207.5911}{{\tt arXiv:1207.5911}}].

\bibitem{Chacko:2012sy}
Z.~Chacko and R.~K. Mishra,  Phys. Rev. {\bf D87} (2013), no.~11 115006,
  [\href{http://xxx.lanl.gov/abs/1209.3022}{{\tt arXiv:1209.3022}}].

\bibitem{Chacko:2012vm}
Z.~Chacko, R.~Franceschini, and R.~K. Mishra,  JHEP {\bf 04} (2013) 015,
  [\href{http://xxx.lanl.gov/abs/1209.3259}{{\tt arXiv:1209.3259}}].

\bibitem{Bellazzini:2012vz}
B.~Bellazzini, C.~Csaki, J.~Hubisz, J.~Serra, and J.~Terning,  Eur. Phys. J.
  {\bf C73} (2013), no.~2 2333, [\href{http://xxx.lanl.gov/abs/1209.3299}{{\tt
  arXiv:1209.3299}}].

\bibitem{Hernandez-Leon:2017kea}
P.~Hernandez-Leon and L.~Merlo,  Phys. Rev. {\bf D96} (2017), no.~7 075008,
  [\href{http://xxx.lanl.gov/abs/1703.02064}{{\tt arXiv:1703.02064}}].

\bibitem{Feruglio:1992wf}
F.~Feruglio,  Int. J. Mod. Phys. {\bf A8} (1993) 4937--4972,
  [\href{http://xxx.lanl.gov/abs/hep-ph/9301281}{{\tt hep-ph/9301281}}].

\bibitem{Contino:2010mh}
R.~Contino, C.~Grojean, M.~Moretti, F.~Piccinini, and R.~Rattazzi,  JHEP {\bf
  05} (2010) 089, [\href{http://xxx.lanl.gov/abs/1002.1011}{{\tt
  arXiv:1002.1011}}].

\bibitem{Alonso:2012px}
R.~Alonso, M.~B. Gavela, L.~Merlo, S.~Rigolin, and J.~Yepes,  Phys. Lett. {\bf
  B722} (2013) 330--335, [\href{http://xxx.lanl.gov/abs/1212.3305}{{\tt
  arXiv:1212.3305}}]. [Erratum: Phys. Lett.B726,926(2013)].

\bibitem{Alonso:2012pz}
R.~Alonso, M.~B. Gavela, L.~Merlo, S.~Rigolin, and J.~Yepes,  Phys. Rev. {\bf
  D87} (2013), no.~5 055019, [\href{http://xxx.lanl.gov/abs/1212.3307}{{\tt
  arXiv:1212.3307}}].

\bibitem{Buchalla:2013rka}
G.~Buchalla, O.~Catà, and C.~Krause,  Nucl. Phys. {\bf B880} (2014) 552--573,
  [\href{http://xxx.lanl.gov/abs/1307.5017}{{\tt arXiv:1307.5017}}]. [Erratum:
  Nucl. Phys.B913,475(2016)].

\bibitem{Brivio:2013pma}
I.~Brivio, T.~Corbett, O.~J.~P. Éboli, M.~B. Gavela, J.~Gonzalez-Fraile, M.~C.
  Gonzalez-Garcia, L.~Merlo, and S.~Rigolin,  JHEP {\bf 03} (2014) 024,
  [\href{http://xxx.lanl.gov/abs/1311.1823}{{\tt arXiv:1311.1823}}].

\bibitem{Brivio:2014pfa}
I.~Brivio, O.~J.~P. Éboli, M.~B. Gavela, M.~C. Gonzalez-Garcia, L.~Merlo, and
  S.~Rigolin,  JHEP {\bf 12} (2014) 004,
  [\href{http://xxx.lanl.gov/abs/1405.5412}{{\tt arXiv:1405.5412}}].

\bibitem{Gavela:2014vra}
M.~B. Gavela, J.~Gonzalez-Fraile, M.~C. Gonzalez-Garcia, L.~Merlo, S.~Rigolin,
  and J.~Yepes,  JHEP {\bf 10} (2014) 044,
  [\href{http://xxx.lanl.gov/abs/1406.6367}{{\tt arXiv:1406.6367}}].

\bibitem{Gavela:2014uta}
M.~B. Gavela, K.~Kanshin, P.~A.~N. Machado, and S.~Saa,  JHEP {\bf 03} (2015)
  043, [\href{http://xxx.lanl.gov/abs/1409.1571}{{\tt arXiv:1409.1571}}].

\bibitem{Brivio:2015kia}
I.~Brivio, M.~B. Gavela, L.~Merlo, K.~Mimasu, J.~M. No, R.~del Rey, and
  V.~Sanz,  JHEP {\bf 04} (2016) 141,
  [\href{http://xxx.lanl.gov/abs/1511.01099}{{\tt arXiv:1511.01099}}].

\bibitem{Gavela:2016bzc}
B.~M. Gavela, E.~E. Jenkins, A.~V. Manohar, and L.~Merlo,  Eur. Phys. J. {\bf
  C76} (2016), no.~9 485, [\href{http://xxx.lanl.gov/abs/1601.07551}{{\tt
  arXiv:1601.07551}}].

\bibitem{Alonso:2016btr}
R.~Alonso, E.~E. Jenkins, and A.~V. Manohar,  Phys. Lett. {\bf B756} (2016)
  358--364, [\href{http://xxx.lanl.gov/abs/1602.00706}{{\tt
  arXiv:1602.00706}}].

\bibitem{Eboli:2016kko}
O.~J.~P. Éboli and M.~C. Gonzalez–Garcia,  Phys. Rev. {\bf D93} (2016),
  no.~9 093013, [\href{http://xxx.lanl.gov/abs/1604.03555}{{\tt
  arXiv:1604.03555}}].

\bibitem{Brivio:2016fzo}
I.~Brivio, J.~Gonzalez-Fraile, M.~C. Gonzalez-Garcia, and L.~Merlo,  Eur. Phys.
  J. {\bf C76} (2016), no.~7 416,
  [\href{http://xxx.lanl.gov/abs/1604.06801}{{\tt arXiv:1604.06801}}].

\bibitem{Alonso:2016oah}
R.~Alonso, E.~E. Jenkins, and A.~V. Manohar,  JHEP {\bf 08} (2016) 101,
  [\href{http://xxx.lanl.gov/abs/1605.03602}{{\tt arXiv:1605.03602}}].

\bibitem{deFlorian:2016spz}
{\bf LHC Higgs Cross Section Working Group} Collaboration, D.~de~Florian {\em
  et.~al.},  \href{http://xxx.lanl.gov/abs/1610.07922}{{\tt arXiv:1610.07922}}.

\bibitem{Merlo:2016prs}
L.~Merlo, S.~Saa, and M.~Sacristán-Barbero,  Eur. Phys. J. {\bf C77} (2017),
  no.~3 185, [\href{http://xxx.lanl.gov/abs/1612.04832}{{\tt
  arXiv:1612.04832}}].

\bibitem{Kozow:2019txg}
P.~Kozów, L.~Merlo, S.~Pokorski, and M.~Szleper,  JHEP {\bf 07} (2019) 021,
  [\href{http://xxx.lanl.gov/abs/1905.03354}{{\tt arXiv:1905.03354}}].

\bibitem{Merlo:2017sun}
L.~Merlo, F.~Pobbe, and S.~Rigolin,  Eur. Phys. J. {\bf C78} (2018), no.~5 415,
  [\href{http://xxx.lanl.gov/abs/1710.10500}{{\tt arXiv:1710.10500}}].

\bibitem{Alonso-Gonzalez:2018vpc}
J.~Alonso-González, L.~Merlo, F.~Pobbe, S.~Rigolin, and O.~Sumensari,
  \href{http://xxx.lanl.gov/abs/1807.08643}{{\tt arXiv:1807.08643}}.

\bibitem{Giudice:2012zp}
G.~F. Giudice, R.~Rattazzi, and A.~Strumia,  Phys. Lett. {\bf B715} (2012)
  142--148, [\href{http://xxx.lanl.gov/abs/1204.5465}{{\tt arXiv:1204.5465}}].

\bibitem{Redi:2012ad}
M.~Redi and A.~Strumia,  JHEP {\bf 11} (2012) 103,
  [\href{http://xxx.lanl.gov/abs/1208.6013}{{\tt arXiv:1208.6013}}].

\bibitem{Redi:2016esr}
M.~Redi and R.~Sato,  JHEP {\bf 05} (2016) 104,
  [\href{http://xxx.lanl.gov/abs/1602.05427}{{\tt arXiv:1602.05427}}].

\bibitem{DiLuzio:2016sbl}
L.~Di~Luzio, F.~Mescia, and E.~Nardi,  Phys. Rev. Lett. {\bf 118} (2017), no.~3
  031801, [\href{http://xxx.lanl.gov/abs/1610.07593}{{\tt arXiv:1610.07593}}].

\bibitem{Farina:2016tgd}
M.~Farina, D.~Pappadopulo, F.~Rompineve, and A.~Tesi,  JHEP {\bf 01} (2017)
  095, [\href{http://xxx.lanl.gov/abs/1611.09855}{{\tt arXiv:1611.09855}}].

\bibitem{Coy:2017yex}
R.~Coy, M.~Frigerio, and M.~Ibe,
  \href{http://xxx.lanl.gov/abs/1706.04529}{{\tt arXiv:1706.04529}}.

\bibitem{Choi:1986zw}
K.~Choi, K.~Kang, and J.~E. Kim,  Phys. Lett. {\bf B181} (1986) 145--149.

\bibitem{Bjorken:1988as}
J.~D. Bjorken, S.~Ecklund, W.~R. Nelson, A.~Abashian, C.~Church, B.~Lu, L.~W.
  Mo, T.~A. Nunamaker, and P.~Rassmann,  Phys. Rev. {\bf D38} (1988) 3375.

\bibitem{Carena:1988kr}
M.~Carena and R.~D. Peccei,  Phys. Rev. {\bf D40} (1989) 652.

\bibitem{Raffelt:2006cw}
G.~G. Raffelt,  Lect. Notes Phys. {\bf 741} (2008) 51--71,
  [\href{http://xxx.lanl.gov/abs/hep-ph/0611350}{{\tt hep-ph/0611350}}].
  [,51(2006)].

\bibitem{Adler:2008zza}
{\bf E787, E949} Collaboration, S.~Adler {\em et.~al.},  Phys. Rev. {\bf D77}
  (2008) 052003, [\href{http://xxx.lanl.gov/abs/0709.1000}{{\tt
  arXiv:0709.1000}}].

\bibitem{Bellini:2012kz}
{\bf Borexino} Collaboration, G.~Bellini {\em et.~al.},  Phys. Rev. {\bf D85}
  (2012) 092003, [\href{http://xxx.lanl.gov/abs/1203.6258}{{\tt
  arXiv:1203.6258}}].

\bibitem{Friedland:2012hj}
A.~Friedland, M.~Giannotti, and M.~Wise,  Phys. Rev. Lett. {\bf 110} (2013),
  no.~6 061101, [\href{http://xxx.lanl.gov/abs/1210.1271}{{\tt
  arXiv:1210.1271}}].

\bibitem{Lees:2013kla}
{\bf BaBar} Collaboration, J.~P. Lees {\em et.~al.},  Phys. Rev. {\bf D87}
  (2013), no.~11 112005, [\href{http://xxx.lanl.gov/abs/1303.7465}{{\tt
  arXiv:1303.7465}}].

\bibitem{Armengaud:2013rta}
E.~Armengaud {\em et.~al.},  JCAP {\bf 1311} (2013) 067,
  [\href{http://xxx.lanl.gov/abs/1307.1488}{{\tt arXiv:1307.1488}}].

\bibitem{Clarke:2013aya}
J.~D. Clarke, R.~Foot, and R.~R. Volkas,  JHEP {\bf 02} (2014) 123,
  [\href{http://xxx.lanl.gov/abs/1310.8042}{{\tt arXiv:1310.8042}}].

\bibitem{Viaux:2013lha}
N.~Viaux, M.~Catelan, P.~B. Stetson, G.~Raffelt, J.~Redondo, A.~A.~R. Valcarce,
  and A.~Weiss,  Phys. Rev. Lett. {\bf 111} (2013) 231301,
  [\href{http://xxx.lanl.gov/abs/1311.1669}{{\tt arXiv:1311.1669}}].

\bibitem{Aprile:2014eoa}
{\bf XENON100} Collaboration, E.~Aprile {\em et.~al.},  Phys. Rev. {\bf D90}
  (2014), no.~6 062009, [\href{http://xxx.lanl.gov/abs/1404.1455}{{\tt
  arXiv:1404.1455}}]. [Erratum: Phys. Rev.D95,no.2,029904(2017)].

\bibitem{Ayala:2014pea}
A.~Ayala, I.~Domínguez, M.~Giannotti, A.~Mirizzi, and O.~Straniero,  Phys.
  Rev. Lett. {\bf 113} (2014), no.~19 191302,
  [\href{http://xxx.lanl.gov/abs/1406.6053}{{\tt arXiv:1406.6053}}].

\bibitem{Khachatryan:2014rra}
{\bf CMS} Collaboration, V.~Khachatryan {\em et.~al.},  Eur. Phys. J. {\bf C75}
  (2015), no.~5 235, [\href{http://xxx.lanl.gov/abs/1408.3583}{{\tt
  arXiv:1408.3583}}].

\bibitem{Mimasu:2014nea}
K.~Mimasu and V.~Sanz,  JHEP {\bf 06} (2015) 173,
  [\href{http://xxx.lanl.gov/abs/1409.4792}{{\tt arXiv:1409.4792}}].

\bibitem{Dolan:2014ska}
M.~J. Dolan, F.~Kahlhoefer, C.~McCabe, and K.~Schmidt-Hoberg,  JHEP {\bf 03}
  (2015) 171, [\href{http://xxx.lanl.gov/abs/1412.5174}{{\tt
  arXiv:1412.5174}}]. [Erratum: JHEP07,103(2015)].

\bibitem{Millea:2015qra}
M.~Millea, L.~Knox, and B.~Fields,  Phys. Rev. {\bf D92} (2015), no.~2 023010,
  [\href{http://xxx.lanl.gov/abs/1501.04097}{{\tt arXiv:1501.04097}}].

\bibitem{Vinyoles:2015aba}
N.~Vinyoles, A.~Serenelli, F.~L. Villante, S.~Basu, J.~Redondo, and J.~Isern,
  JCAP {\bf 1510} (2015), no.~10 015,
  [\href{http://xxx.lanl.gov/abs/1501.01639}{{\tt arXiv:1501.01639}}].

\bibitem{Aad:2015zva}
{\bf ATLAS} Collaboration, G.~Aad {\em et.~al.},  Eur. Phys. J. {\bf C75}
  (2015), no.~7 299, [\href{http://xxx.lanl.gov/abs/1502.01518}{{\tt
  arXiv:1502.01518}}]. [Erratum: Eur. Phys. J.C75,no.9,408(2015)].

\bibitem{Krnjaic:2015mbs}
G.~Krnjaic,  Phys. Rev. {\bf D94} (2016), no.~7 073009,
  [\href{http://xxx.lanl.gov/abs/1512.04119}{{\tt arXiv:1512.04119}}].

\bibitem{Marciano:2016yhf}
W.~J. Marciano, A.~Masiero, P.~Paradisi, and M.~Passera,  Phys. Rev. {\bf D94}
  (2016), no.~11 115033, [\href{http://xxx.lanl.gov/abs/1607.01022}{{\tt
  arXiv:1607.01022}}].

\bibitem{Izaguirre:2016dfi}
E.~Izaguirre, T.~Lin, and B.~Shuve,  Phys. Rev. Lett. {\bf 118} (2017), no.~11
  111802, [\href{http://xxx.lanl.gov/abs/1611.09355}{{\tt arXiv:1611.09355}}].

\bibitem{Brivio:2017ije}
I.~Brivio, M.~B. Gavela, L.~Merlo, K.~Mimasu, J.~M. No, R.~del Rey, and
  V.~Sanz,  Eur. Phys. J. {\bf C77} (2017), no.~8 572,
  [\href{http://xxx.lanl.gov/abs/1701.05379}{{\tt arXiv:1701.05379}}].

\bibitem{Bauer:2017nlg}
M.~Bauer, M.~Neubert, and A.~Thamm,  Phys. Rev. Lett. {\bf 119} (2017), no.~3
  031802, [\href{http://xxx.lanl.gov/abs/1704.08207}{{\tt arXiv:1704.08207}}].

\bibitem{Anastassopoulos:2017ftl}
{\bf CAST} Collaboration, V.~Anastassopoulos {\em et.~al.},  Nature Phys. {\bf
  13} (2017) 584--590, [\href{http://xxx.lanl.gov/abs/1705.02290}{{\tt
  arXiv:1705.02290}}].

\bibitem{Dolan:2017osp}
M.~J. Dolan, T.~Ferber, C.~Hearty, F.~Kahlhoefer, and K.~Schmidt-Hoberg,  JHEP
  {\bf 12} (2017) 094, [\href{http://xxx.lanl.gov/abs/1709.00009}{{\tt
  arXiv:1709.00009}}].

\bibitem{Jaeckel:2015jla}
J.~Jaeckel and M.~Spannowsky,  Phys. Lett. {\bf B753} (2016) 482--487,
  [\href{http://xxx.lanl.gov/abs/1509.00476}{{\tt arXiv:1509.00476}}].

\bibitem{Bauer:2017ris}
M.~Bauer, M.~Neubert, and A.~Thamm,  JHEP {\bf 12} (2017) 044,
  [\href{http://xxx.lanl.gov/abs/1708.00443}{{\tt arXiv:1708.00443}}].

\bibitem{Acciarri:1994gb}
{\bf L3} Collaboration, M.~Acciarri {\em et.~al.},  Phys. Lett. {\bf B345}
  (1995) 609--616.

\bibitem{Anashkin:1999da}
{\bf DELPHI} Collaboration, E.~Anashkin {\em et.~al.},  in {\em {Proceedings,
  International Europhysics Conference on High Energy Physics (Eps-Hep 1999):
  Tampere, Finland, July 15-21, 1999}}, 1999.

\bibitem{Aubert:2008as}
{\bf BaBar} Collaboration, B.~Aubert {\em et.~al.},  in {\em {Proceedings, 34th
  International Conference on High Energy Physics (ICHEP 2008): Philadelphia,
  Pennsylvania, July 30-August 5, 2008}}, 2008.
\newblock \href{http://xxx.lanl.gov/abs/0808.0017}{{\tt arXiv:0808.0017}}.

\bibitem{delAmoSanchez:2010ac}
{\bf BaBar} Collaboration, P.~del Amo~Sanchez {\em et.~al.},  Phys. Rev. Lett.
  {\bf 107} (2011) 021804, [\href{http://xxx.lanl.gov/abs/1007.4646}{{\tt
  arXiv:1007.4646}}].

\bibitem{Seong:2018gut}
{\bf Belle} Collaboration, I.~S. Seong {\em et.~al.},
  \href{http://xxx.lanl.gov/abs/1809.05222}{{\tt arXiv:1809.05222}}.

\bibitem{Merlo:2019anv}
L.~Merlo, F.~Pobbe, S.~Rigolin, and O.~Sumensari,  JHEP {\bf 06} (2019) 091,
  [\href{http://xxx.lanl.gov/abs/1905.03259}{{\tt arXiv:1905.03259}}].

\bibitem{Coan:1997ye}
{\bf CLEO} Collaboration, T.~E. Coan {\em et.~al.},  Phys. Rev. Lett. {\bf 80}
  (1998) 1150--1155, [\href{http://xxx.lanl.gov/abs/hep-ex/9710028}{{\tt
  hep-ex/9710028}}].

\bibitem{Bardeen:1978nq}
W.~A. Bardeen, S.~H.~H. Tye, and J.~A.~M. Vermaseren,  Phys. Lett. {\bf 76B}
  (1978) 580--584.

\bibitem{DiVecchia:1980yfw}
P.~Di~Vecchia and G.~Veneziano,  Nucl. Phys. {\bf B171} (1980) 253--272.

\bibitem{Barr:1992qq}
S.~M. Barr and D.~Seckel,  Phys. Rev. {\bf D46} (1992) 539--549.

\bibitem{Kamionkowski:1992mf}
M.~Kamionkowski and J.~March-Russell,  Phys. Lett. {\bf B282} (1992) 137--141,
  [\href{http://xxx.lanl.gov/abs/hep-th/9202003}{{\tt hep-th/9202003}}].

\bibitem{Holman:1992us}
R.~Holman, S.~D.~H. Hsu, T.~W. Kephart, E.~W. Kolb, R.~Watkins, and L.~M.
  Widrow,  Phys. Lett. {\bf B282} (1992) 132--136,
  [\href{http://xxx.lanl.gov/abs/hep-ph/9203206}{{\tt hep-ph/9203206}}].

\bibitem{Alonso:2017avz}
R.~Alonso and A.~Urbano,  \href{http://xxx.lanl.gov/abs/1706.07415}{{\tt
  arXiv:1706.07415}}.

\end{thebibliography}

\providecommand{\href}[2]{#2}\begingroup\raggedright\endgroup

\end{document}